\documentclass[]{mn2e}
\usepackage{epsfig}
\usepackage{graphicx}
\usepackage{mathtools}
\usepackage{pgf}

\title[M\,33 monitoring. V]{The UK Infrared Telescope M\,33 monitoring
project. V. The star formation history across the galactic disc}
\author[Javadi et al.]{\parbox{16cm}{
                       Atefeh Javadi$^{1}$,
                       Jacco Th.\ van Loon$^{2}$,
                       Habib G. Khosroshahi$^{1}$,\\
                       Fatemeh Tabatabaei$^{3,4}$,
                       Roya Hamedani Golshan$^{1,5}$
                       and
                       Maryam Rashidi$^{6}$\vspace{2mm}}\\
$^{1}$School of Astronomy, Institute for Research in Fundamental Sciences
      (IPM), P.O.\ Box 19395-5531, Tehran, Iran\\
$^{2}$Lennard-Jones Laboratories, Keele University, ST5 5BG, UK\\
$^{3}$Instituto de Astrof\'{\i}sica de Canarias, V\'{\i}a L\'actea S/N,
      E-38205 La Laguna, Spain\\
$^{4}$Departamento de Astrof\'{\i}sica, Universidad de La Laguna, E-38206 La
      Laguna, Spain\\
$^{5}$Department of Physics, Faxculty of Science, University of Isfahan,
      Isfahan, P.O.\ Box 81746-73441, Iran\\
$^{6}$Physics Department, Alzahra University, Vanak, 1993891176, Tehran, Iran}
\date{Submitted: 2016}
\pagerange{\pageref{firstpage}--\pageref{lastpage}}
\pubyear{2016}
\begin{document}
\maketitle
\label{firstpage}
\begin{abstract}
We have conducted a near-infrared monitoring campaign at the UK InfraRed
Telescope (UKIRT), of the Local Group spiral galaxy M\,33 (Triangulum). On the
basis of their variability, we have identified stars in the very final stage
of their evolution, and for which the luminosity is more directly related to
the birth mass than the more numerous less-evolved giant stars that continue
to increase in luminosity. In this fifth paper of the series, we construct the
birth mass function and hence derive the star formation history across the
galactic disc of M\,33. The star formation rate has varied between
$\sim0.010\pm0.001$ ($\sim0.012\pm0.007$) and 0.060$\pm0.005$ (0.052$\pm0.009$)
   M$_\odot$ yr$^{-1}$ kpc$^{-2}$ statistically (systematically)
in the central square kiloparsec of
M\,33, comparable with the values derived previously with another camera. The
total star formation rate in M\,33 within a galactocentric radius of 14 kpc
has varied between
$\sim0.110\pm0.005$ ($\sim0.174\pm0.060$) and $\sim0.560\pm0.028$ ($\sim0.503\pm0.100$)
  M$_\odot$ yr$^{-1}$ statistically (systematically).
We find evidence of two
epochs during which the star formation rate was enhanced by a factor of a few
-- one that started $\sim6$ Gyr ago and lasted $\sim3$ Gyr and produced $\geq71$\%
of the total mass in stars, and one $\sim250$ Myr ago that lasted $\sim200$ Myr
and formed $\leq13$\% of the mass in stars. Radial star formation history profiles
suggest that the inner disc of M\,33 was formed in an inside--out formation scenario.
The outskirts of the disc are dominated by the old population, which may be the 
result of dynamical effects over many Gyr. We find correspondence to spiral structure
for all stars, but enhanced only for stars younger than $\sim100$ Myr; this suggests
that the spiral arms are transient features and not part of a global density wave
potential.
\end{abstract}
\begin{keywords}stars: luminosity function, mass function
-- galaxies: evolution
-- galaxies: individual: M\,33
-- galaxies: spiral
-- galaxies: stellar content
-- galaxies: structure
\end{keywords}
%========================================================================== 1
\section{Introduction}

Galactic evolution is driven at the end-points of stellar evolution, where
copious mass loss returns chemically-enriched and sometimes dusty matter back
to the interstellar medium (ISM); the stellar winds of evolved stars and the
violent deaths of the most massive stars also inject energy and momentum into
the ISM, generating turbulence and galactic fountains when superbubbles pop as
they reach the ``surface'' of the galactic disc. Evolved stars are also
excellent tracers, not just of the feedback processes, but also of the
underlying populations, that were formed from millions to billions of years
prior to their appearance. The evolved phases of evolution generally represent
the most luminous, and often the coolest, making evolved stars brilliant
beacons at IR wavelengths, where it is also easier to see them deep inside
galaxies as dust is more transparent at those longer wavelengths than in the
optical and ultraviolet where their progenitors shine. The final stages of
evolution of stars with birth masses up to $M\sim30$ M$_\odot$ -- Asymptotic
Giant Branch (AGB) stars and red supergiants (RSGs) -- are characterised by
strong radial pulsations of the cool atmospheric layers, rendering them
identifiable as long-period variables (LPVs) in photometric monitoring
campaigns spanning months to years (e.g., Whitelock, Feast \& Catchpole 1991;
Wood et al.\ 1992; Wood 2000; Ita et al.\ 2004a,b).

The prevalent theory of disc galaxy formation hinges on the formation of
isolated discs through the dissipational collapse of a gaseous proto-galaxy
embedded in cold dark matter halos (White \& Rees 1978; Fall \& Efstathiou
1980; Peebles 1984). Thin discs are built through gas accretion, while
spheroids and thick discs are built via mergers (Steinmetz \& Navarro 2002;
Brook et al.\ 2004; Brooks et al.\ 2009). An alternative scenario for disc
formation is based on gas-rich major mergers (Robertson et al.\ 2006; Stewart
et al.\ 2009). In fact, each of these scenarios may operate depending on
galaxy environments (Blanton et al.\ 2003). Spatially-resolved observational
estimates of the star formation histories (SFHs) -- and stellar chemical
abundances -- in spiral galaxies are a robust way to constrain models for
their formation and evolution (Eggen et al.\ 1962; Freeman \& Bland-Hawthorn
2002). This has become possible within galaxies of the Local Group (e.g.,
Sarajedini et al.\ 2000; Harbeck et al.\ 2001; Ferguson et al.\ 2002; Brown et
al.\ 2003; Cole et al.\ 2005). All those methods rely on theoretical models of
stellar evolution to provide stellar ages and evolutionary lifetimes, as well
as stellar yields, and an assumption about the initial mass function (IMF) at
which stars are born.

M\,33 is a low luminosity, late type disc spiral galaxy in the Local Group of
galaxies. With a stellar mass of $M_\star\sim3$--$6\times10^9$ M$_\odot$ and
neutral gas mass of $M_{\rm gas}\sim3\times10^9$ M$_\odot$ (Corbelli 2003),
M\,33 is less massive but more gas-rich than the Milky Way disc, which has
$M_\star\sim5\times10^{10}$ M$_\odot$ (Licquia \& Newman; cf.\ McMillan 2011)
and $M_{\rm gas}\sim1\times10^{10}$ M$_\odot$ (Nakanishi \& Sofue 2015), with no
sign of recent mergers. In addition to its proximity (distance modulus
$\mu=24.9$ mag; Bonanos et al.\ 2006; U et al.\ 2009), its large -- but not
too large -- angular size ($\sim1^\circ$) and low inclination ($56^\circ$)
offer us a unique opportunity to study its stellar populations, their history
and their feedback across an entire spiral galaxy. Using colour--magnitude
diagrams (CMDs) of resolved stars in M\,33, Barker et al.\ (2007) found a
positive age gradient in three fields just outside the break radius on the
southern minor axis. More recently, Williams et al.\ (2009) found a negative
age gradient within the break radius in the radial mass profile, $r<6$ kpc, in
four fields on M\,33’s southern major axis. This hints at multiple mechanisms
at play, that shaped the M\,33 disc as we see it today, and calls for
independent, homogeneous analyses across the entire span of the M\,33 galaxy.
Here we provide such study.

The main objectives of our project are described in Javadi, van Loon \&
Mirtorabi (2011c): to construct the mass function of LPVs and derive from this
the SFH in M\,33; to correlate spatial distributions of the LPVs of different
mass with galactic structures (spheroid, disc and spiral arm components); to
measure the rate at which dust is produced and fed into the ISM; to establish
correlations between the dust production rate, luminosity, and amplitude of an
LPV; and to compare the {\it in situ} dust replenishment with the amount of
pre-existing dust. Paper I in the series presented the photometric catalogue
of stars in the inner square kpc (Javadi et al.\ 2011a), with Paper II
presenting the galactic structure and SFH (Javadi, van Loon and Mirtorabi
2011b), and Paper III presenting the mass-loss mechanism and dust production
rate (Javadi et al.\ 2013). Paper IV presented the photometric catalogue of
stars in a nearly square degree area covering much of the M\,33 optical disc.
This fifth paper in the series covers the SFH across the enlarged area; the
next paper will present an analysis of the mass return to the M\,33 disc.

%========================================================================== 2
\section{Input data and models}
%-------------------------------------------------------------------------- 2.1
\subsection{Catalogue of variable stars}

In paper IV we described the method employed to identify large-amplitude LPVs
across the galactic disc of M\,33 with the WFCAM imager on the United Kingdom
IR Telescope (UKIRT) on Mauna Kea, Hawai'i. The area that was monitored covers
almost a square degree ($53^\prime\times53^\prime$); given the inclination and
distance of M\,33, we thus probe all of its disc out to a galactocentric
radius of 7.5 kpc, i.e.\ 0.88 $R_{25}$ (optical radius), and out to 15 kpc
(1.76 $R_{25}$) in certain directions. The observations were done between
2005--2007 in the K$_{\rm s}$-band ($\lambda=2.2$ $\mu$m) with occasionally
observations in J- and H-bands ($\lambda=1.28$ and 1.68 $\mu$m, respectively)
for the purpose of obtaining colour information.

% FIGURE 1
\begin{figure}
\centerline{\psfig{figure=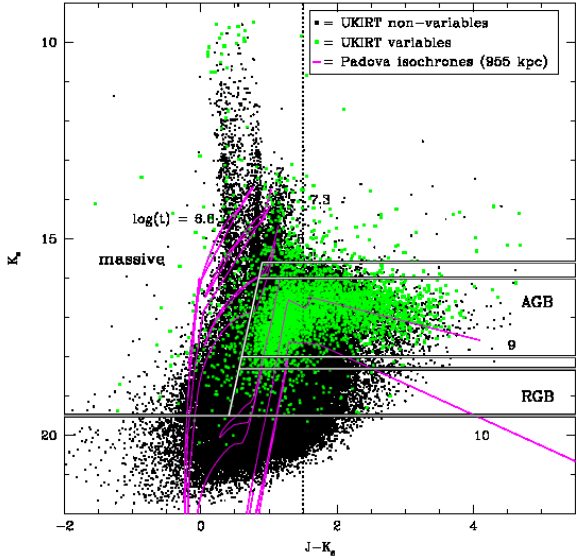,width=84mm}}
\caption[]{K$_{\rm s}$ versus $(J-K_{\rm s})$, with WFCAM variable stars in
green. Overplotted are isochrones from Marigo et al.\ (2008) for solar
metallicity and a distance modulus of $\mu=24.9$ mag,
labelled by their logarithmic ages -- these include circumstellar
     reddening, showing up as large excursions towards red colours;
     carbonaceous dust (e.g., at $\log t=9$) and oxygenous dust (e.g., at
     $\log t=10$) have different reddening slopes. The vertical dashed line at
     $(J-K_{\rm s})=1.5$ mag indicates the colour criterion redwards of which a
     correction is applied for this reddening. The thick black-and-white lines
     demarcate the bulk of the populations of massive, AGB and RGB stars
     (separated by small buffers in K-band magnitude to avoid
     cross-contamination).}

\end{figure}

The photometric catalogue comprises 403\,734 stars, among which 4643 stars were
identified as LPVs -- AGB stars, super-AGB stars and RSGs. Their distribution
over magnitude and colour is shown in Fig.\ 1. 
Some brighter variable RSGs are found (around $K_{\rm s}\sim14$ mag), 
  but the clump of variable stars with $K_{\rm s}<11$ mag  are foreground
  stars and all are saturated. These stars are removed from further analysis.
The amplitude is generally A$_K\leq1$ mag, but a small fraction of
variable stars (8\%) reach A$_K>2$ mag. Among these extreme variables,
311 have six or fewer measurements and their amplitudes are therefore
unreliable. Disregarding those, only 20 stars remain with A$_K>3$ mag.

% FIGURE 2
\begin{figure}
\centerline{\psfig{figure=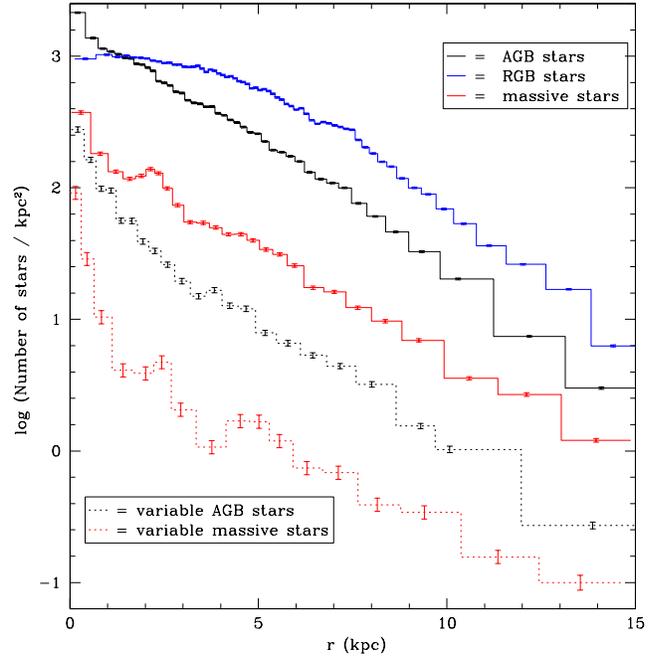,width=84mm}}
\caption[]{Radial distributions across a large part of M\,33 of the near-IR
populations of cool evolved stars, deprojected onto the M\,33 galactic plane.
Dashed lines are for the variable stars.}
\end{figure}

     We separate the stars in our catalogue into three categories: massive
     stars, AGB stars and RGB stars, on the basis of K$_{\rm s}$-band magnitude
     and $(J-K_{\rm s})$ colour criteria (see Fig.\ 1). We define a demarcation
     line between hot massive stars and cooler, less-massive giant stars to
     run from $(J-K_{\rm s},K_{\rm s})=(0.6,18)$ to
     $(J-K_{\rm s},K_{\rm s})=(0.9,15.6)$ mag, with massive stars those that
     have $(J-K_{\rm s}$ colours bluer than this (down to $K_{\rm s}=19.5$ mag)
     or that have $K_{\rm s}<15.6$ mag, and AGB stars and RGB stars those that
     have $(J-K_{\rm s})$ colours redder than this and that have
     $16<K_{\rm s}<18$ mag (AGB) or $18.3<K_{\rm s}<19.5$ mag (RGB, which does
     extend to much fainter magnitudes than our survey completeness). We left
     small gaps to avoid contamination.
The distributions of different types of near-IR populations across the
galactic disc of M\,33 is shown in Fig.\ 2, deprojected onto the M\,33
galactic plane, with dashed lines showing the distributions for the variables
that were identified in our near-IR monitoring programme. There is a clear
difference between the stellar populations within the inner $r<1$ kpc, with
RGB stars showing a core but massive stars -- in particular those identified
by their variability -- showing a cusp. Further out, all populations show a
distribution consistent with an exponential disc. An exponential profile fit
of the form $I=I_0\exp(-r/r_0)$ yields $r_0=2.3$ for the AGB stars ($r_0=2.6$
kpc for the variables) and $r_0=2.6$ kpc for the massive stars ($r_0=3.4$ kpc
for the variables, with a Sersi\'c core). The RGB population, dominated by old
stars, is fit with a Sersi\'c profile, with $r_0=5$ kpc. This inflated
distribution could be a sign of dynamical relaxation. The relatively large
scalelength of the massive stars, which still reflect their birth sites,
compared to the AGB stars may be an indication of the inside--out formation of
the disc of M\,33, whilst the central hike in density of the massive stars may
indicate accretion-induced star formation (see Paper II).

%-------------------------------------------------------------------------- 2.2
\subsection{Stellar evolution models}

The method used here to derive the SFH strongly depends on the AGB phase of
stellar evolution. To date, the Padova models (Marigo et al.\ 2008) are the
most realistic. The full AGB evolution is calculated -- except for the
super-AGB phase of stellar evolution ($M_{\rm birth}\sim6$--9 M$_\odot$: Siess
2007; Doherty et al.\ 2015), which is too fast to be calculated without a
separate treatment. It accurately accounts for the third dredge-up, and hot
bottom burning (HBB; Iben \& Renzini 1983) in the most massive AGB stars
($M_{\rm birth}\sim4$--6 M$_\odot$). They also include the effects of molecular
opacities in the cool stellar mantles and atmospheres, which becomes
especially apparent in the transformation from oxygen-dominated (M-type) AGB
stars to carbon stars in the birth mass range $M_{\rm birth}\sim1.5$--4
M$_\odot$ (Girardi \& Marigo 2007). In addition, and crucial for our method,
the Padova models predict the pulsation properties and the effects of
circumstellar dust production on the spectral energy distribution (SED) of the
star plus its dust envelope. Separate models cover the massive stars, of which
the lower mass range ($M_{\rm birth}\sim9$--30 M$_\odot$ become RSGs. The full
suite of models thus enables us to evaluate the SFH from as long ago as 10
Gyr, to as recent as $\sim10$ Myr ago.

That said, those models do not consider the effects of magnetic fields or
rotation on the mixing process inside the stars. Including more efficient
mixing would make stars live longer and massive stars evolve to cooler RSGs
(Meynet \& Maeder 2000; Heger, Woosley \& Spruit 2005). Likewise, the life
times of intermediate-mass stars are prolonged in the presence of convective
overshoot (Marigo \& Giardi 2007). Also, the thermal pulsing AGB is computed
by adopting parameterisations for the efficiency of third dredge-up, and the
threshold temperature at which it sets in, not by solving the equations
governing the stellar internal structure.

%========================================================================== 3
\section{Methodology: deriving the SFH}

In paper II, we presented a novel way to derive the SFH by using cool,
large-amplitude LPVs which are identified in IR monitoring programmes. We use
the fact that these variables have reached the very final stages of their
evolution, and their brightness can thus be transformed into their mass at
birth by employing theoretical evolutionary tracks, or by isochrones. The LPVs
are located at the cool end of each of the isochrones. Hence, we convert the
observed K-band magnitude (which is close to the peak in the SED, thus
minimising uncertainties in the bolometric corrections) into the birth mass.

This was done in paper II for four different metallicities, from super-solar
-- suitable for massive elliptical galaxies and stellar populations in the
bulges of massive spiral galaxies such as the Milky Way -- to sub-solar
values appropriate for galaxies such as the Magellanic Clouds (Rezaeikh et
al.\ 2014). As discussed in paper II, the central region of M\,33 has an
approximately solar metallicity (Rosolowsky \& Simon 2008; Magrini et al.\
2009), for which we adopt $Z=0.015$. The disc of M\,33 outside that central
region is characterised by sub-solar metallicity; in paper IV we showed that
$Z=0.008$ agrees well with our CMDs as well as the literature. Magrini et
al.\ (2007) derived an oxygen abundance gradient based on H\,{\sc ii} regions
of $-0.19$ dex kpc$^{-1}$ for $r<3$ kpc and $-0.038$ dex kpc$^{-1}$ for $r\geq3$
kpc, suggesting a possibly slightly larger range of metallicities. However, in
Paper II we showed that the SFHs derived using $Z=0.0008$ and $Z=0.015$ are
very similar and hence we are not worried about small deviations from these
typical values.

Since the method was described in detail in paper II, here we only provide a
brief summary. The total mass of the stars created between times $t$ and
$t+{\rm d}t$  is defined as:
\begin{equation}
{\rm d}M(t)=\xi(t)\,{\rm d}t,
\end{equation}
where $\xi(t)$ is the star formation rate (SFR), in M$_\odot$ yr$^{-1}$. The
mass $M(t)$ corresponds to the number of stars formed, $N$, as:
\begin{equation}
{\rm d}N(t)={\rm d}M(t)\frac{\int_{\rm min}^{\rm max}f_{\rm IMF}(m)\,{\rm d}m}
{\int_{\rm min}^{\rm max}f_{\rm IMF}(m)m\,{\rm d}m},
\end{equation}
where  $f_{\rm IMF}$ is the IMF, which is described by:
\begin{equation}
f_{\rm IMF}=Am^{-\alpha},
\end{equation}
where $A$ is the normalisation factor and $\alpha$ is defined for different
ranges in stellar mass, $m$, by Kroupa (2001):
\begin{equation}
\alpha=\left\{
\begin{array}{lll}
+0.3\pm0.7 & {\rm for} & {\rm min}\leq m/{\rm M}_\odot<0.08 \\
+1.3\pm0.5 & {\rm for} & 0.08\leq m/{\rm M}_\odot<     0.50 \\
+2.3\pm0.3 & {\rm for} & 0.50\leq m/{\rm M}_\odot<{\rm max} \\
\end{array}
\right.
\end{equation}
For the minimum and maximum of the stellar mass ranges we adopt
${\rm min}=0.02$ M$_\odot$ and ${\rm max}=200$ M$_\odot$, respectively.

The question now is how many of these stars, $n$, are LPVs around time $t$,
which is when we observe them. The number of stars with mass between $m(t)$
and $m(t+{\rm d}t)$ that were created between times $t$ and $t+{\rm d}t$ is:
\begin{equation}
{\rm d}n(t)={\rm d}N(t)\frac{\int_{m(t)}^{m(t+dt)}f_{\rm IMF}(m)\,{\rm d}m}
{\int_{\rm min}^{\rm max}f_{\rm IMF}(m)\,{\rm d}m}.
\end{equation}
Combining the above equations we have:
\begin{equation}
{\rm d}n(t)=\xi(t){\rm d}t\
\frac{\int_{m(t)}^{m(t+{\rm d}t)}f_{\rm IMF}(m){\rm d}m}
{\int_{\rm min}^{\rm max}f_{\rm IMF}(m)m\,{\rm d}m}.
\end{equation}
When we determine $\xi(t)$ over an age bin ${\rm d}t$, the number of LPVs
observed in that age bin, ${\rm d}n^\prime$, depends on the duration of the
evolutionary phase of large-amplitude, long-period variability, the
``pulsation duration'' $\delta t$:
\begin{equation}
{\rm d}n^\prime(t)={\rm d}n(t)\frac{\delta t}{{\rm d}t}.
\end{equation}
Inverting the above equations, we obtain a relation between the observed
number of LPVs in a certain age bin, and the SFR that long ago:
\begin{equation}
\xi(t)=\frac{{\rm d}n^\prime(t)}{\delta t}\
\frac{\int_{\rm min}^{\rm max}f_{\rm IMF}(m)m\,{\rm d}m}
{\int_{m(t)}^{m(t+{\rm d}t)}f_{\rm IMF}(m)\,{\rm d}m}.
\end{equation}
Note that the value of the normalisation constant $A$ in Eq.\ (3) does not
matter.

The relation between K-band magnitude and mass was constructed in paper II
from the points on each of the Padova isochrones when stars reach their peak
brightness in the K-band. The relation was monotonic for AGB stars and RSGs;
in the mass regime $0.7<\log(M/{\rm M}_\odot)<1.2$--1.3 the Padova isochrones
suggest an excursion towards fainter K-band magnitudes. The stellar evolution
models for these super-AGB stars are truncated. We compared two different
approaches to deal with this: [1] interpolation of the $M$--$K$ relation
through the mass range of super-AGB stars, or [2] accepting the jump to
fainter magnitudes from AGB to super-AGB stars. The second approach yielded
unrealistic results for the SFH while the first approach was credible, so we
adopt the interpolation approach here also.

Dusty LPVs suffer from attenuation. While the isochrones of Marigo et al.\
(2008) include an estimate of the photometric effects of circumstellar dust
their values are very uncertain. Instead, we correct for the effects of
reddening by tracing the reddened stars, along the reddening vector, back to
the point on the isochrone where they would lie in the absence of reddening.
As Fig.\ 1 illustrates, there exist two types of reddening; carbon stars
(e.g., at $t=1$ Gyr, $\log t=9$) redden very rapidly, resulting in a
relatively shallow trajectory in the CMD, whereas oxygen-rich stars (such as
low-mass stars at $t=10$ Gyr, $\log t=10$) experience a ``greyer'' type of
reddening due to the greater transparency of silicates. We thus obtain
reddening-free K-band magnitudes, $K_0$, from the relation:
\begin{equation}
K_0=K+a(1.25-(J-K)),
\end{equation}
with $a=0.52$ mag mag$^{-1}$ for carbon stars and $a=0.72$ mag mag$^{-1}$ for
oxygen-rich stars at $Z=0.015$, and $a=0.64$ mag mag$^{-1}$ for carbon stars
and $a=0.78$ mag mag$^{-1}$ for oxygen-rich stars at $Z=0.008$. The reddening
correction will be applied to stars with $(J-K)>1.5$ mag. In the case where no
J-band data are available we use a similar reddening correction prescribed as
a function of $(H-K)$ colour. In the case neither J- nor H-band data are
available we assume a detection limit of $J=21$ mag.

It is therefore important to distinguish between carbon stars and oxygen-rich
stars. Unfortunately, such information is not available for the majority of
our LPVs in M\,33. Instead, we fix the mass range for carbon stars; based on
theory and observation in the LMC (Groenewegen \& de Jong 1993; van Loon,
Marshall \& Zijlstra 2005; Girardi \& Marigo 2007) we expect at solar and
slightly sub-solar metallicity carbon stars to descend from stars in the
range $M_{\rm birth}\sim1.5$--4 M$_\odot$ -- at low mass the third dredge-up is
not strong enough to change the composition of the surface to carbon
dominated, while HBB prevents carbon to enrich the surface in higher mass
stars. In practice, we first apply the carbon dust correction to all reddened
stars; if the derived mass is in the 1.5--4 M$_\odot$ range then we accept that
star as a carbon star, otherwise we re-apply an oxygenous dust correction.

We refer to Paper II for the relations converting the de-reddened K-band
magnitude into birth mass, and the birth mass into LPV age, and for the
parameterisation of the pulsation duration. To have a meaningful errorbar on
the SFR in each age bin, instead of using equal age bins we adjust the sizes
of the age bins such that each contain the same number of stars. Despite the
rarity of massive LPVs born relatively recently, we are still more able to
detect hikes in the star formation rate within the past Gyr than 10 Gyr ago.
So a recent ``bursty'' SFH does not preclude this from having been a feature
throughout the assembly of the galaxy.

Finally, in Paper III we realised that the duration of the LPV phase that we
detect in our survey is overestimated in the Padova models, resulting in too
low values for the SFR. We thus applied a correction to the pulsation
duration, and hence to the SFRs that had been derived in Paper II. We will
revisit this important issue in detail for the disc and central region in the
discussion section.

     Furthermore, we stress that, while our method is model dependent, the
     Padova models are the only models that predict the pulsation behaviour,
     which is essential information for our method to work. We would like to
     encourage other groups to also predict the pulsation behaviour, to offer
     another avenue to empirically explore the differences between models.
     Other models that computate the thermal-pulsing AGB evolution (but, like
     the Padova models, not for super-AGB stars) include BaSTI (Pietrinferni
     et al.\ 2004, 2006). In the Appendix we compare the Mass--Luminosity and
     Mass--Age relations between the BaSTI and Padova models. The main
     difference is that the faintest stars, around $\sim17.6$ mag) are less
     massive and older when using the BaSTI models ($\sim12$ Gyr compared to
     10 Gyr when using the Padova models), but the difference is negligible
     for intermediate-age stars (few Gyr old).

%========================================================================== 4
\section{Results}
%-------------------------------------------------------------------------- 4.1
\subsection{The SFH in the central region of M\,33}

Since we already derived the SFH for the central region of M\,33 using UIST
data in paper II, we first show how this compares to the SFH derived from the
WFCAM data.

     The distribution over brightness and the present day (birth) mass
     function of LPVs are shown in Fig.\ 3.
Two peaks can be seen, one at $\log(M/{\rm M}_\odot)\sim0.1$ and another at
$\log(M/{\rm M}_\odot)\sim0.5$. As described in paper IV, the branch of stars
brighter than $K_{\rm s}< 11$ mag was removed from further analysis because
these stars are affected by saturation. However, RSGs in M\,33 are fainter
than this, so the derived mass function is not affected.

% FIGURE 3
\begin{figure}
\centerline{\vbox{
\psfig{figure=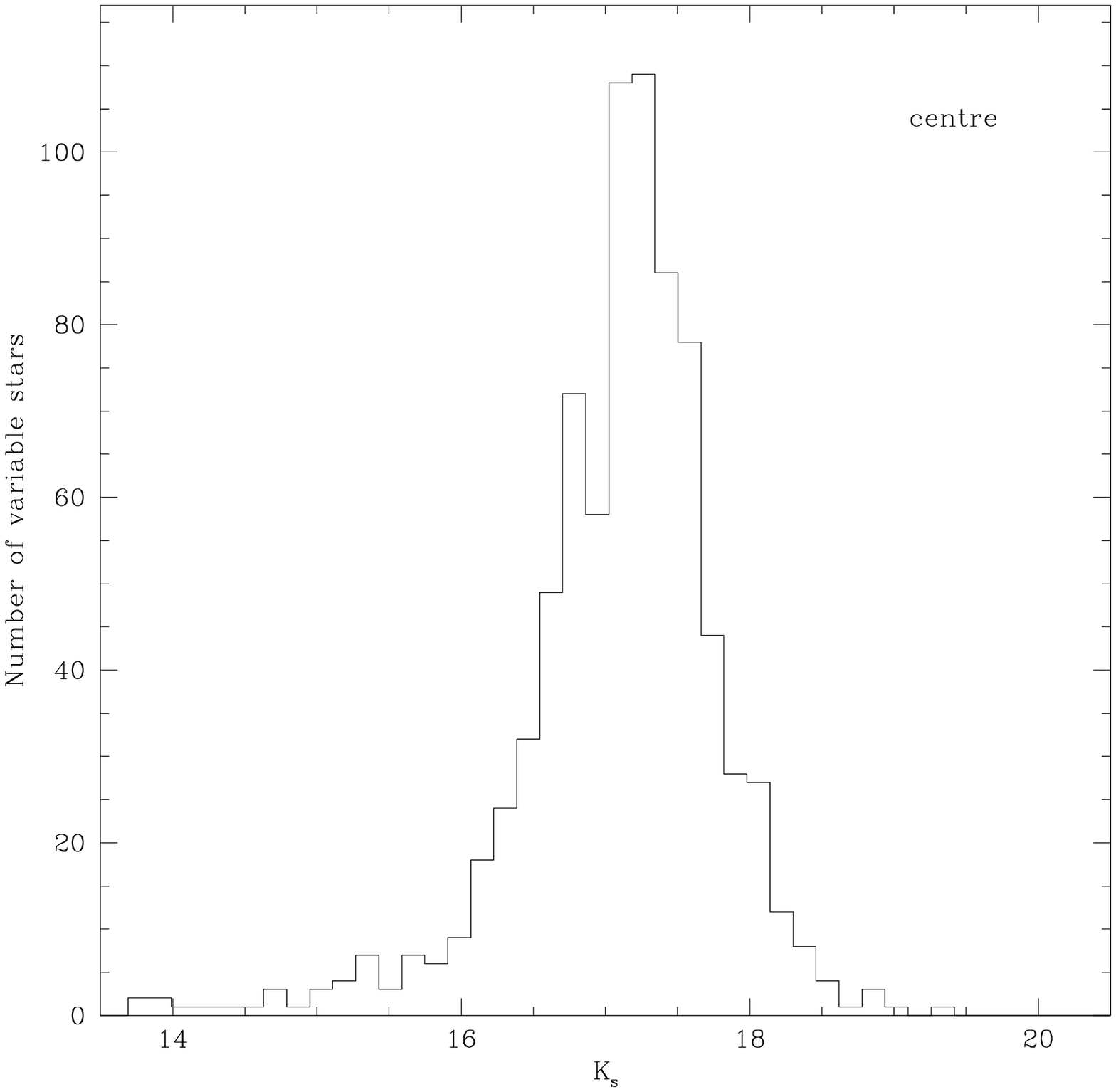,width=84mm}
\psfig{figure=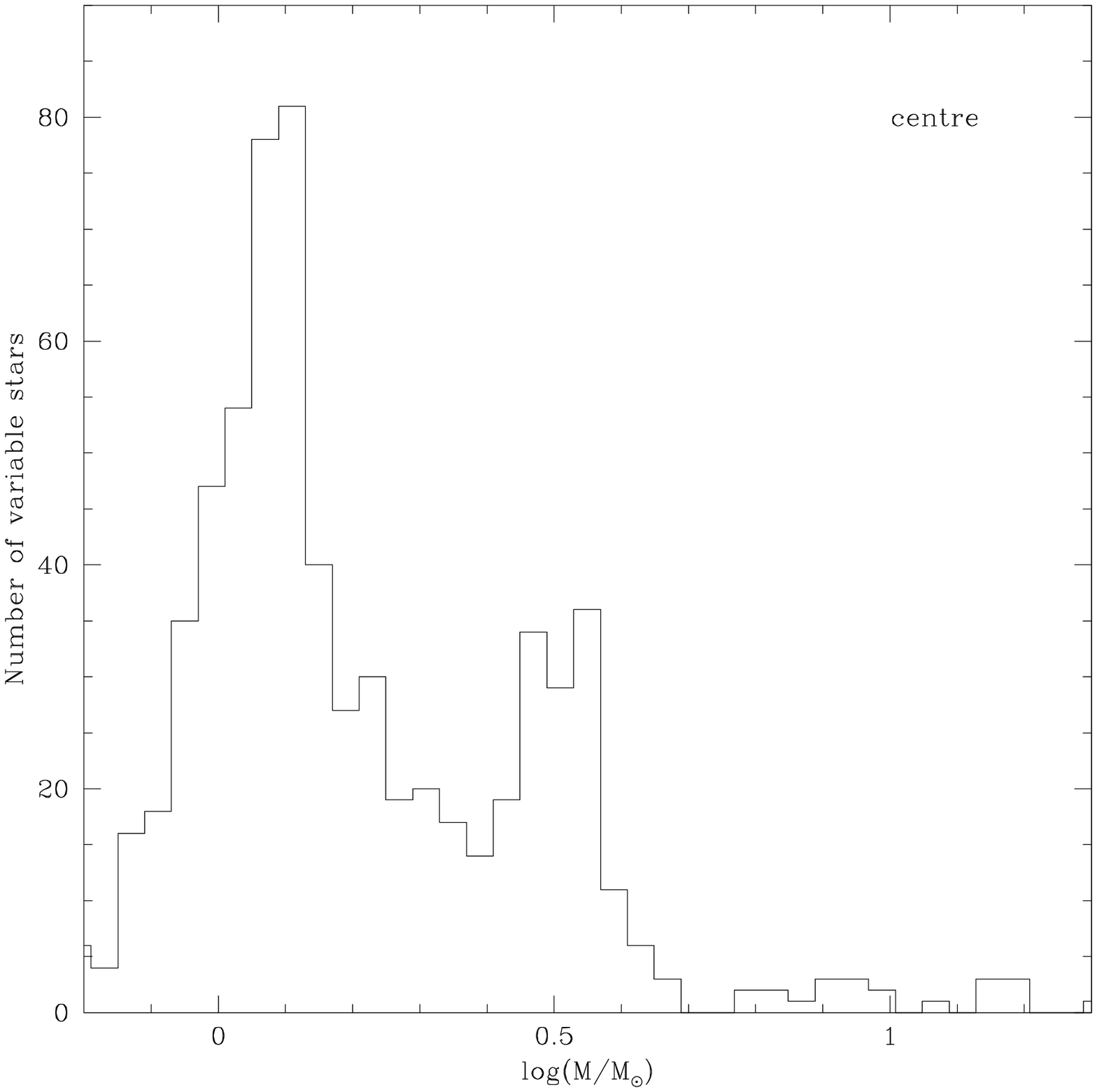,width=84mm}
}}
\caption[]{
     Top: distribution of large-amplitude variable stars, as a function of
     near-IR brightness in the central square kpc of M\,33. Bottom: the
     derived present-day mass function.
}
\end{figure}

% FIGURE 4
\begin{figure}
\centerline{\psfig{figure=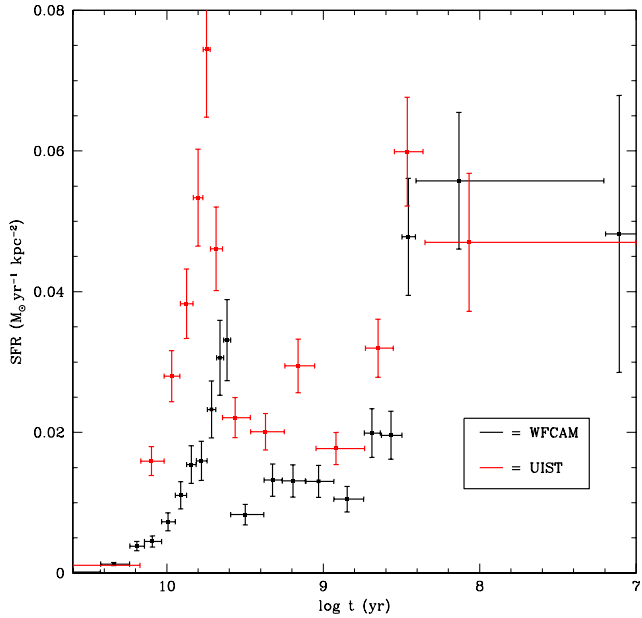,width=84mm}}
\caption[]{The SFH in the central square kpc of M\,33 derived from pulsating
AGB stars and RSGs using ({\it black:}) WFCAM data and ({\it red:}) UIST data;
    these rates include a correction of a factor of 7 (WFCAM) or 10 (UIST) --
     see text.
}
\end{figure}

The SFH of the central square kpc of M\,33 is shown in Fig.\ 4, separately as
derived from the WFCAM data and from the UIST data.
     The oldest bin is only shown in part as it extends to (unrealistically)
     large ages. The only reason it is plotted is to show that, as expected,
     the star formation rate we estimate for ages exceeding the Hubble time is
     negligible.
Two main epochs of star
formation are obvious; a major epoch of formation $\approx4$--5 Gyr ago
($\log t=9.6$--9.7) peaking around 4 Gyr ago at a level $\sim2.5$ times as
high as during the subsequent couple of Gyr. This is slightly shifted to more
recent times compared with the UIST result for which this epoch is
$\approx4$--8 Gyr ago ($\log t=9.6$--9.9) with a peak around 6 Gyr ago ($\log
t=9.8$). This could arise from the inferior angular resolution of WFCAM,
resulting in incompleteness and blends rendering stars brighter (hence
appearing younger) than they really are. The UIST data may be affected too,
if less so (the ancient SFR is higher than in the WFCAM data), and hence the
peak at 4--6 Gyr may not be real and star formation could well have started --
and peaked -- earlier. However, this does not change our final conclusion that
the SFR in the central regions of M\,33 has dropped considerably some 4 Gyr
ago.

A second epoch of star formation is seen to occur from 300 Myr--20 Myr ago
($\log t=8.5$--7.2), in both data sets, with a rate $\sim1.5$ times higher
than the aforementioned peak. There is a hint for the SFR to have been
decreasing slightly over the course of this more recent epoch, suggesting that
something may have happened $\sim300$ Myr ago which triggered an enhanced SFR.

%-------------------------------------------------------------------------- 4.2
\subsection{The SFH in the disc of M\,33}

% FIGURE 5
\begin{figure}
\centerline{\vbox{
\psfig{figure=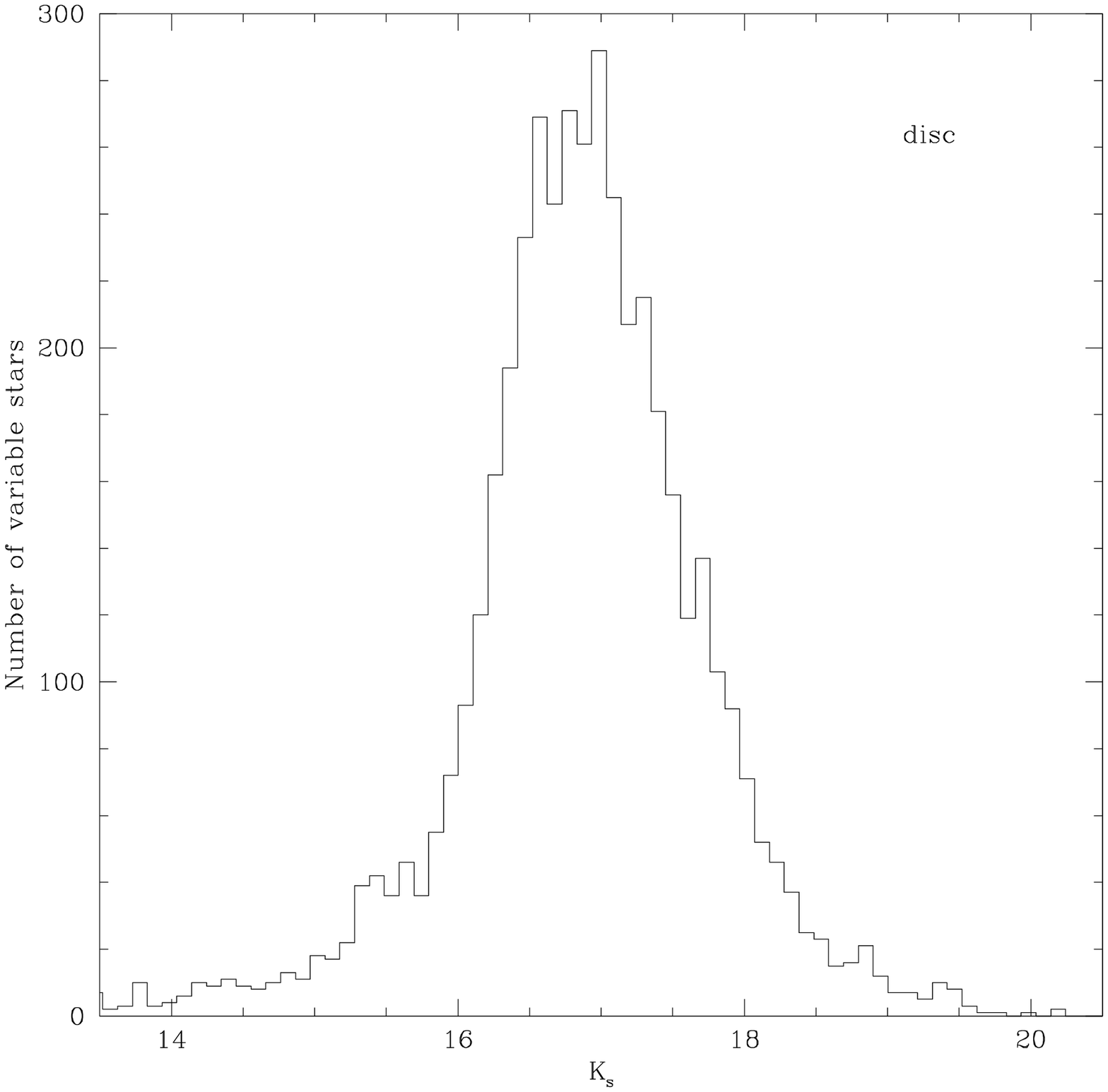, width=84mm}
\psfig{figure=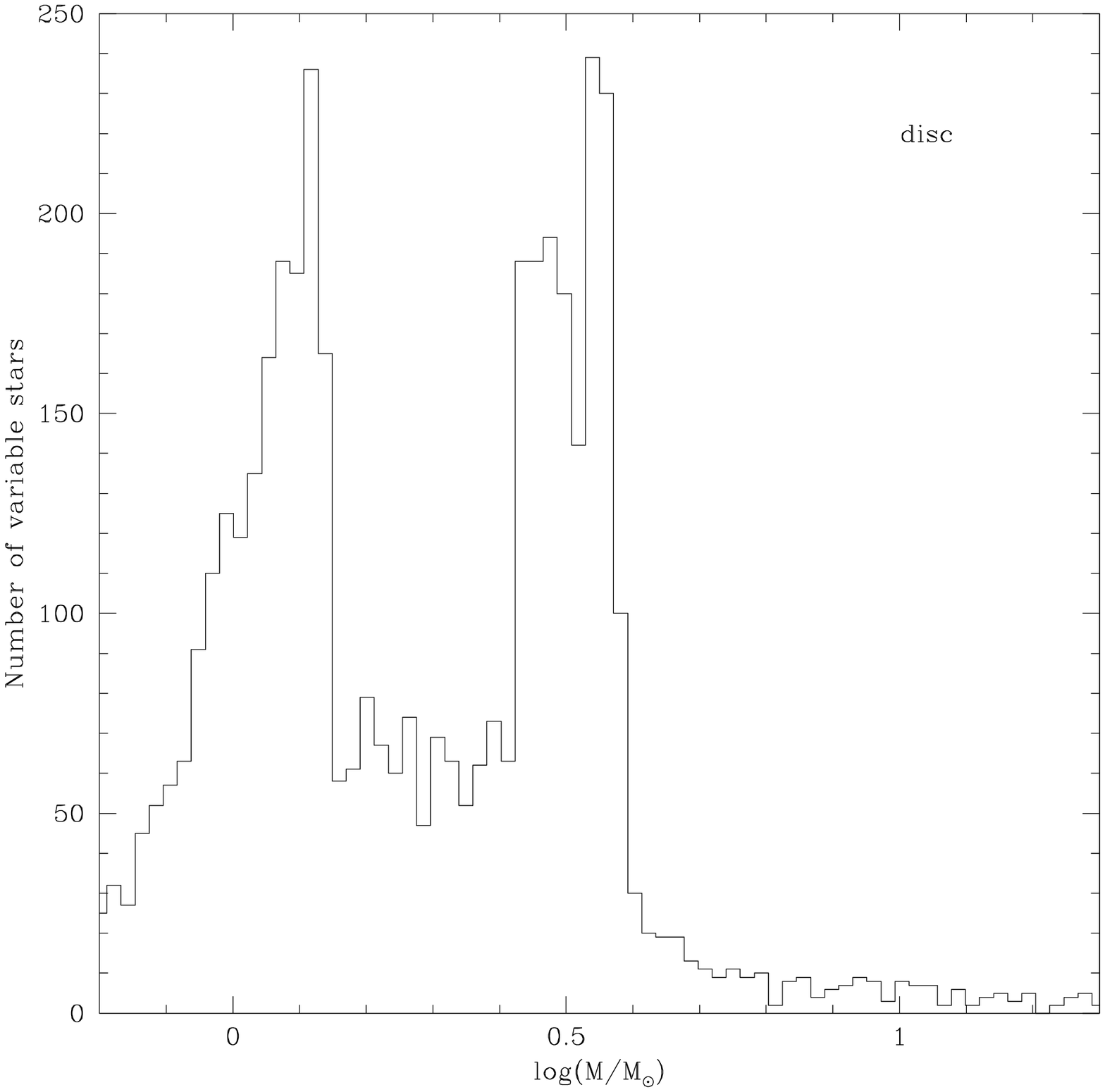, width=84mm}
}}
\caption[]{
     Top: distribution of large-amplitude variable stars, as a function of
     near-IR brightness in the entire disc of M\,33. Bottom: the derived
     present-day mass function.
}
\end{figure}

The WFCAM images reach out to a galactocentric radius of nearly 15 kpc, which
includes the pseudo-bulge and most of the disc and spiral arms of M\,33.
 The distribution over brightness and
the present day (birth) mass function of LPVs in this larger area are shown in
Fig.\ 5. Again, the mass distribution is dominated by two populations of LPVs;
one around $\log(M/{\rm M}_\odot)=0.1$ and another around
$\log(M/{\rm M}_\odot=0.5$. There are relatively many intermediate-age AGB
stars in the disc compared with the central region, and this suggests that the
average age of stars in the disc is lower than in the central region.

% FIGURE 6
\begin{figure}
\centerline{\psfig{figure=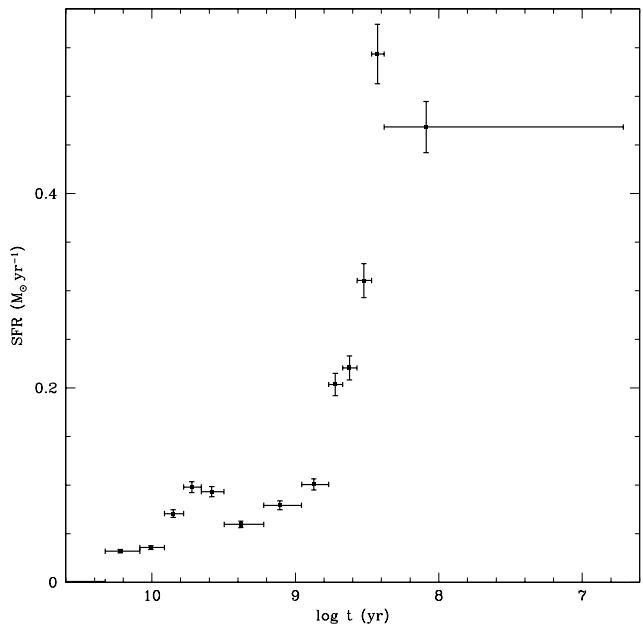,width=84mm}}
\caption[]{The SFH across the entire disc of M\,33.
A correction factor of 7 has been applied to the SFRs (see text).
}
\end{figure}

The SFH across the disc of M\,33 is shown in Fig.\ 6. The double-peaked mass
function is again translated into two epochs of enhanced star formation. The
old epoch took place $\approx3$--6 Gyr ago ($\log t=9.5$--9.8), peaking
$\sim5$ Gyr ago ($\log t=9.7$) at a level almost twice that in the subsequent
few Gyr. Barker et al.\ (2011) showed that the outer disc of M\,33 had a major
epoch of star formation $\sim2$--4 Gyr ago ($z\sim0.2$--0.4 for a standard
WMAP $\Lambda$CDM cosmology; Dunkley et al.\ 2009). The offset by one Gyr
compared to the timing of this epoch of star formation in the central region
(derived from the superior UIST data), suggests that the outer regions of
M\,33 are younger than the central region. This would confirm the results by
Li et al.\ (2004). This conclusion is, however, controversial since Cioni et
al.\ (2008) suggested the opposite. Because their results were based on
photometry with WFCAM, they may have been affected by the crowding in the
central regions.

The second, recent epoch of enhanced star formation occurred $\approx200$--300
Myr ago ($\log t=8.3$--8.5), reaching a level of almost 4 times that during
the earlier epoch of star formation. We caution, though, that this is almost
certainly due to the difference in time resolution: the peak of the earlier
epoch is an average over $\sim1$ Gyr and peaks lasting only $\sim100$ Myr may
have been much more intense than that. What is beyond doubt, however, is that
the more recent peak in SFR is relatively strong compared to the same epoch in
the central region of M\,33.

While our approach may favour the assignment of the status of carbon star over
oxygen-rich star, the SFH results do not change substantially when assuming
that all stars are oxygen-rich (see Appendix).

%-------------------------------------------------------------------------- 4.3
\subsection{Spatial variation of the SFH across the disc}

The variation of SFH as a function of radial distance from the centre of M\,33
($r$) is shown in Fig.\ 7. Each radius bin contains the same number of
variable stars; as a result, the bins grow in size with increasing distance.
Two important results become clear immediately; firstly, the old epoch of star
formation weakens and then practically disappears as one moves outward through
the disc. In the outermost parts of the disc, the star formation rate
gradually increases until $t\sim200$ Myr ($\log t=8.3$) ago, and then
decreases again in more recent times. Secondly, the very recent epoch of star
formation is seen across the entire disc at a similar rate, but lower in the
very central region. More subtly, star formation at intermediate ages,
$t\sim1$ Gyr ($\log t=9$) is most pronounced in the most prominent part of the
disc, $2<r<7$ kpc, but weaker both in the central and the outermost regions.

Gratier et al.\ (2010) showed that the radial profile of total baryonic mass
in M\,33 displays a double exponential component, with a break around $r\sim6$
kpc. Within that radius, while the star formation efficiency is relatively
constant the SFH shifts to more recent times as one moves outward; this is
referred to as the ``inside--out'' formation scenario. Our analysis (Fig.\ 7)
clearly confirms this scenario for the disc of M\,33. Fig.\ 7 also suggests a
change in SFH around $r\sim6$ kpc; beyond that radius, the star formation
efficiency decreases (see also Robles-Valdez et al.\ 2013).

% FIGURE 7
\begin{figure}
\centerline{\psfig{figure=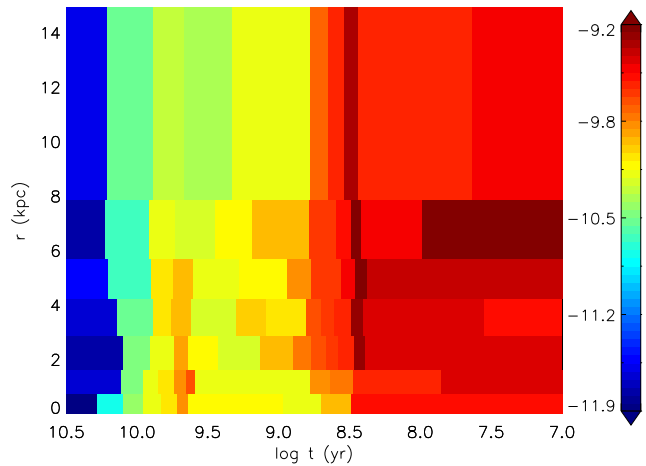,width=84mm}}
\caption[]{SFH at different radial distance from the centre of M\,33, in the
galaxy (de-projected) plane. The colour bar values are on a logarithmic scale
and represent the fraction of the total stellar mass as it formed each year.
The inverse of that is the time it takes at that rate, to form all the stars
that have ever formed. Each radius bin contains the same number of stars.}
\end{figure}

% FIGURE 8
\begin{figure}
\centerline{\psfig{figure=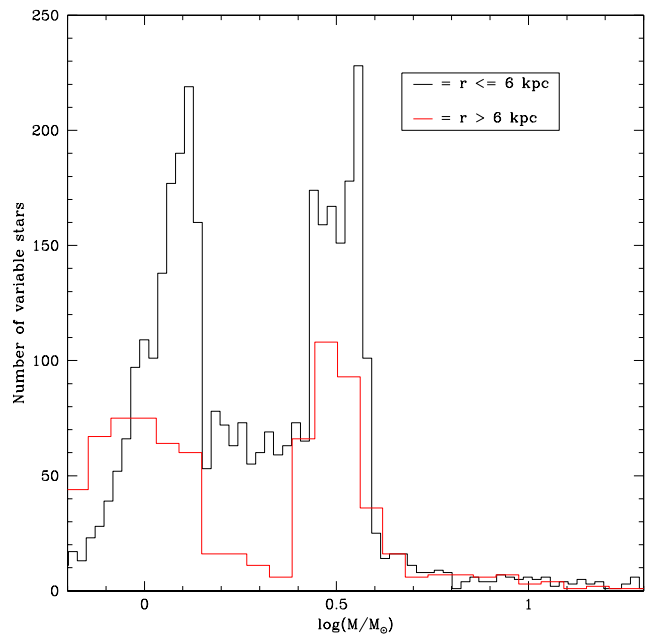,width=84mm}}
\caption[]{Present-day mass function of large-amplitude variable stars at
$r\leq 6$ kpc (black) and at $r>6$ kpc (red).}
\end{figure}

To examine the difference between the disc within $r\leq 6$ kpc and the outer
regions at $r>6$ kpc, the mass functions for these two areas are shown
separately in Fig.\ 8. Apart from an overall drop in stellar content, the
outer regions display a conspicuous dearth of intermediate-mass stars (between
$\log(M/{\rm M}_\odot)\sim0.15$--0.4). On the other hand, the outer regions
contain more -- even in an absolute sense -- low-mass, presumably old LPVs
($\log(M/{\rm M}_\odot)<0$). In these sparsely populated regions, stars at
$\log(M/{\rm M}_\odot)>0.4$ could be foreground stars; we provide an estimate
of this in Section 4.4. The corresponding SFHs are shown in Fig.\ 9. Both at
$r\leq 6$ kpc and at $r>6$ kpc there is strong evidence for recent star
formation. The old epoch of star formation is present in both areas as well.
It is possible that the old stars in the outermost regions were formed further
in, but migrated outwards through dynamical relaxation or interaction with
spiral arms (Grand, Kawata \& Cropper 2012).

% FIGURE 9
\begin{figure}
\centerline{\psfig{figure=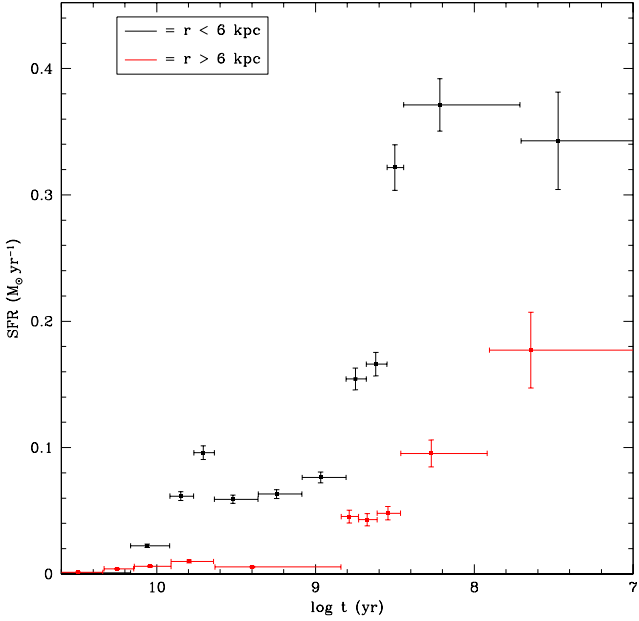,width=84mm}}
\caption[]{The SFH at $r\leq 6$ kpc (black) and at $r>6$ kpc (red); these
rates include a correction of a factor of 7 (see text)
and were derived from near-equally sized areas
.}
\end{figure}

%-------------------------------------------------------------------------- 4.4
\subsection{Spatial distribution of stellar populations}

We have divided the stellar populations into four different age bins, and plot
their spatial distributions in Fig.\ 10. All populations are centrally
concentrated, but this is especially true for the youngest and -- to a lesser
extent -- the oldest. The intermediate-age populations are distributed more
uniformly across the disc. It is interesting to note also the asymmetry in the
distributions, with a higher density towards the south.
     The youngest population exhibits a local overdensity in the North as
     well, around $23\rlap{.}^\circ5$ ($1^{\rm h}34^{\rm m}$) in right ascension
     and $+30\rlap{.}^\circ9$ in declination.
The spiral structure is not obvious in these graphs.

To assess the level of contamination by foreground stars we used the TRILEGAL
simulation tool (Girardi et al.\ 2005). We used the default parameters for the
structure of the Galaxy, simulating a 0.68 square degree field in the
direction of M\,33 ($l=133.61^\circ$, $b=-31.33^\circ$). The region with which
we compare this simulation is the area in the galaxy plane that avoided the
central $r<6$ kpc. The result of this comparison is shown in Fig.\ 11. The
variable stars with $0.3<(J-K_{\rm s})<0.9$ mag experience only a modest level
of contamination ($<8$\%); those with $K_{\rm s}<18.2$ mag and $(J-K_{\rm s})>1$
mag are not affected at all. The level of contamination for all sources with
$(J-K_{\rm s}<0.87$ mag and those with $(J-K_{\rm s})>0.87$ mag and
$K_{\rm s}>18.2$ mag is $<23$\%. These are all upper limits, as most foreground
stars would not appear as LPVs.

% FIGURE 10
\begin{figure*}
\vbox{\hbox{
\psfig{figure=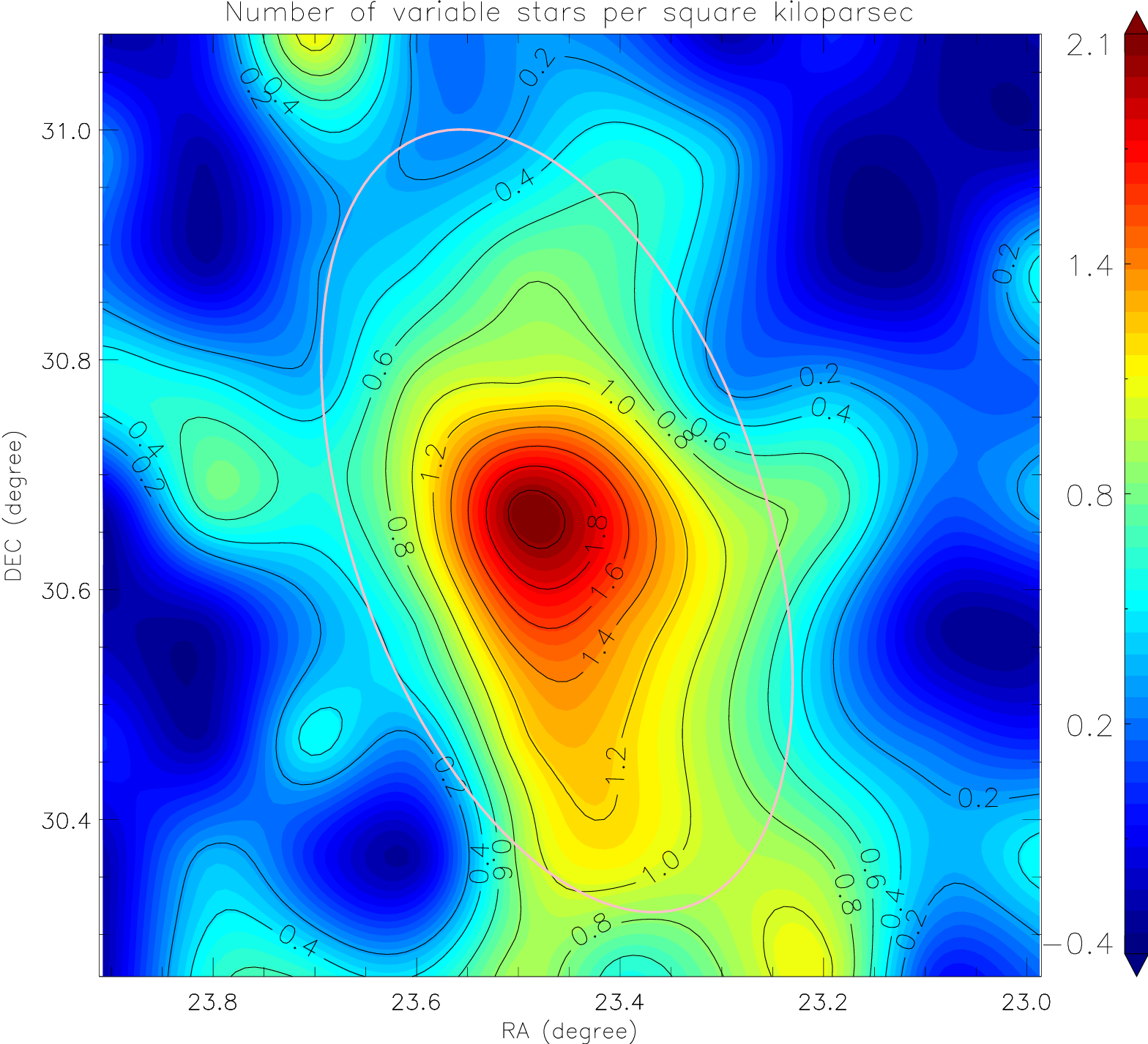,width=85mm}
\hspace*{1mm}
\psfig{figure=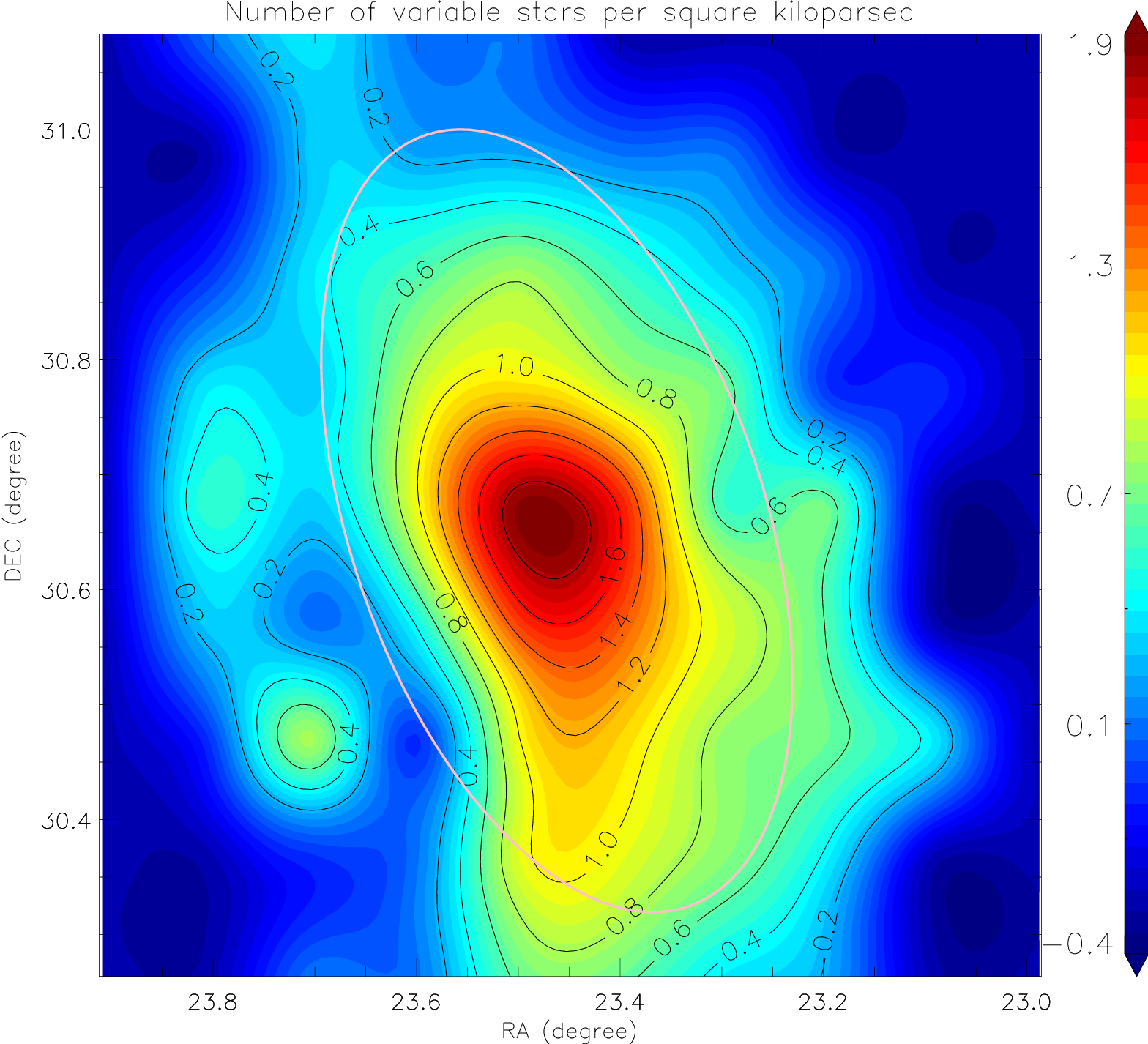,width=85mm}
}
\vspace*{3mm}
\hbox{
\psfig{figure=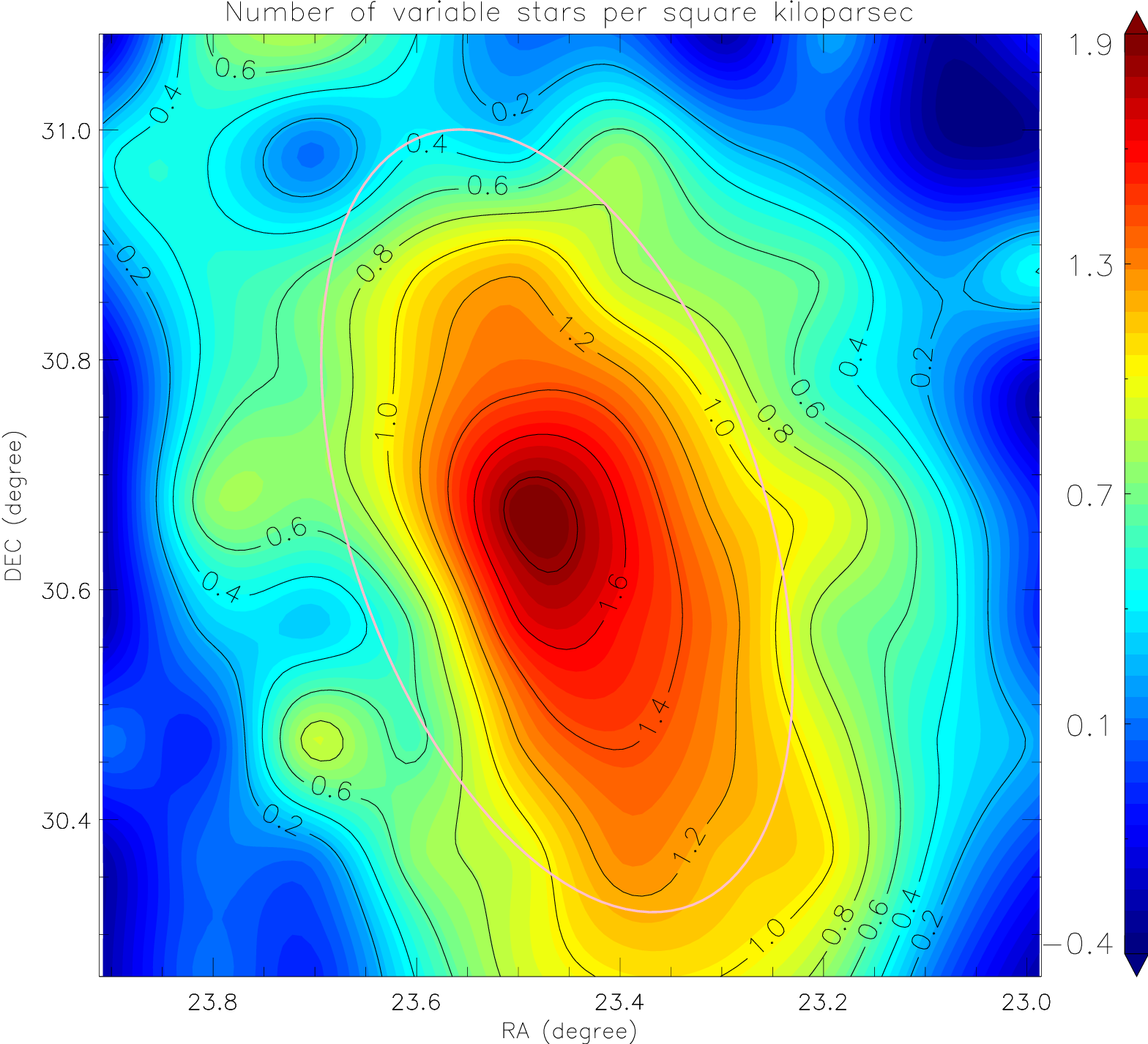,width=85mm}
\hspace*{1mm}
\psfig{figure=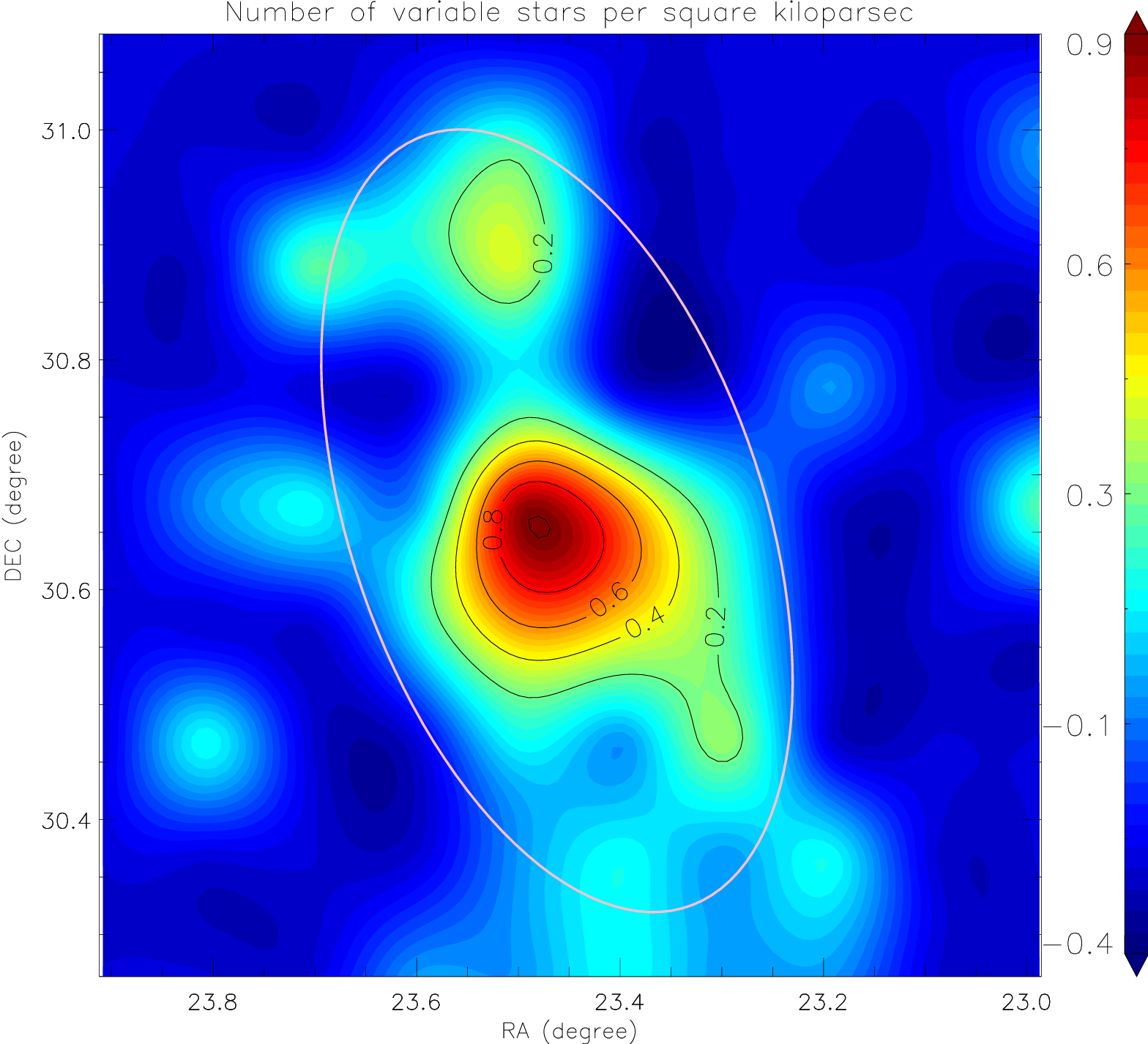,width=85mm}
}}
\caption[]{Spatial distribution across M\,33 of the near-IR LPV populations of
({\it Top left:}) $t>5$ Gyr; ({\it Top right:}) 5 Gyr $<t<1$ Gyr; ({\it Bottom
Left:}) 1 Gyr $<t<100$ Myr; ({\it Bottom right:}) $t<100$ Myr. The units are
logarithmic in number of stars per square kpc. The pink ellipse delineates a
galactocentric radius of 6 kpc.}
\end{figure*}

%-------------------------------------------------------------------------- 4.5
\subsection{SFH and the spiral arm pattern}

M\,33 has two main, inner spiral arms -- I\,N (the northern branch) and I\,S
(the southern branch), indicated by the spiral curves in Fig.\ 12. These
logarithmic arms have a pitch angle of $\theta=27^\circ$ and can be traced out
to $r=3.2$ kpc; they include some of the most prominent H\,{\sc ii} regions in
M\,33: NGC\,604 ($1^{\rm h}34^{\rm m}33\rlap{.}^{\rm s}2$,
$+30^\circ47^\prime06^{\prime\prime}$), IC\,143
($1^{\rm h}34^{\rm m}07\rlap{.}^{\rm s}0$, $+30^\circ47^\prime23^{\prime\prime}$),
IC\,142 ($1^{\rm h}33^{\rm m}55\rlap{.}^{\rm s}8$,
$+30^\circ45^\prime20^{\prime\prime}$), NGC\, 595
($1^{\rm h}33^{\rm m}34\rlap{.}^{\rm s}0$, $+30^\circ41^\prime30^{\prime\prime}$) in
I\,N and IC\,137 ($1^{\rm h}33^{\rm m}29\rlap{.}^{\rm s}3$,
$+30^\circ31^\prime21^{\prime\prime}$) in I\,S. However, the spiral arm pattern
of M\,33 is much more complex, with at least four additional sets of arms that
can be traced out to radii of $r\sim30^\prime$, i.e.\ $>8$ kpc (e.g., Newton
1980; Hagen et al.\ 2015; Sandage \& Humphreys 1980). A sketch of these
spirals based on the location of stellar associations is shown in Fig.\ 12
(Humphreys \& Sandage 1978). Most of the arms are covered completely in our
survey except for a small part of arms IV\,N and V\,N.
    We note previous attempts were made by Block et al.\ (2007) to trace the
     spiral arm structure using near-IR maps of evolved stars; they identified
     major arcs or carbon stars.

Logarithmic spirals have a form $r=r_0\exp(b\theta)$ when seen in a face-on
view. The pitch angle is defined by $u=\tan^{-1}[b^{-1}(\pi/180)\log e]$ where
$b$ is assumed to be constant. Parameters for the ten optical arms in M\,33
have been given in Sandage \& Humphreys (1980), including $b$ and $r_0$ and
the angles (north through east) over which each of the arms exists. To compare
the distributions of stars with these logarithmic arm structures, the
positions of the stars in the image plane must be deprojected onto the galaxy
plane. Then, we can determine the angular distance of stars with respect to
each of the arms.

If the spiral arms are the result of a rigidly rotating density wave, then we
can express the arc distance, $L$, in terms of the time since the passage of
the density wave:
\begin{equation}
t=\frac{L}{r\times[\Omega_{\rm star}-\Omega_{\rm pattern}(r)]}
\end{equation}
We adopt $\Omega_{\rm pattern}=25$ km s$^{-1}$ kpc$^{-1}$ from Newton (1980).
For the stellar velocity function we adopt (Corbelli \& Schneider 1997; in km
s$^{-1}$):
\begin{equation}
v=101.2\tanh\left(\frac{r}{7.2}\right)
+8.5\left[1+\tanh\left(\frac{r-32.6}{8.7}\right)\right],
\end{equation}
where $r$ is the radial distance expressed in arcminutes. This places the
co-rotation radius at $r=3.9$ kpc, beyond which the number of H\,{\sc ii}
regions drops significantly. Obviously, near the co-rotation radius we would
not measure any asymmetric lag between stars just born in the density
enhancement and older stars dying in the disc behind the wave (diffusion
processes would lead to symmetric dispersal).
% Fig.\ 13 illustrates this:
The calculated time lag $t$ of LPVs with respect to arm II\,N becomes
meaningless within a few hundred pc from the co-rotation radius. We should
therefore limit calculations to an area outside the co-rotation radius
% (green lines in Fig.\ 13)
or set a limit on $t$ ($-0.5<t<0.5$ Gyr).
% for the red lines in Fig.\ 13).
The resulting distributions over the time lag, by avoiding an area in the disc
within $\pm0.3$ kpc from the co-rotation radius, is shown in Fig.\ 13. We also
inspected the results we would obtain by placing a limit on $t$, as well as by
adopting a slightly more extended exclusion zone
% (dashed green lines in Fig.\ 13)
-- see Appendix.

% FIGURE 11
\begin{figure}
\centerline{\psfig{figure=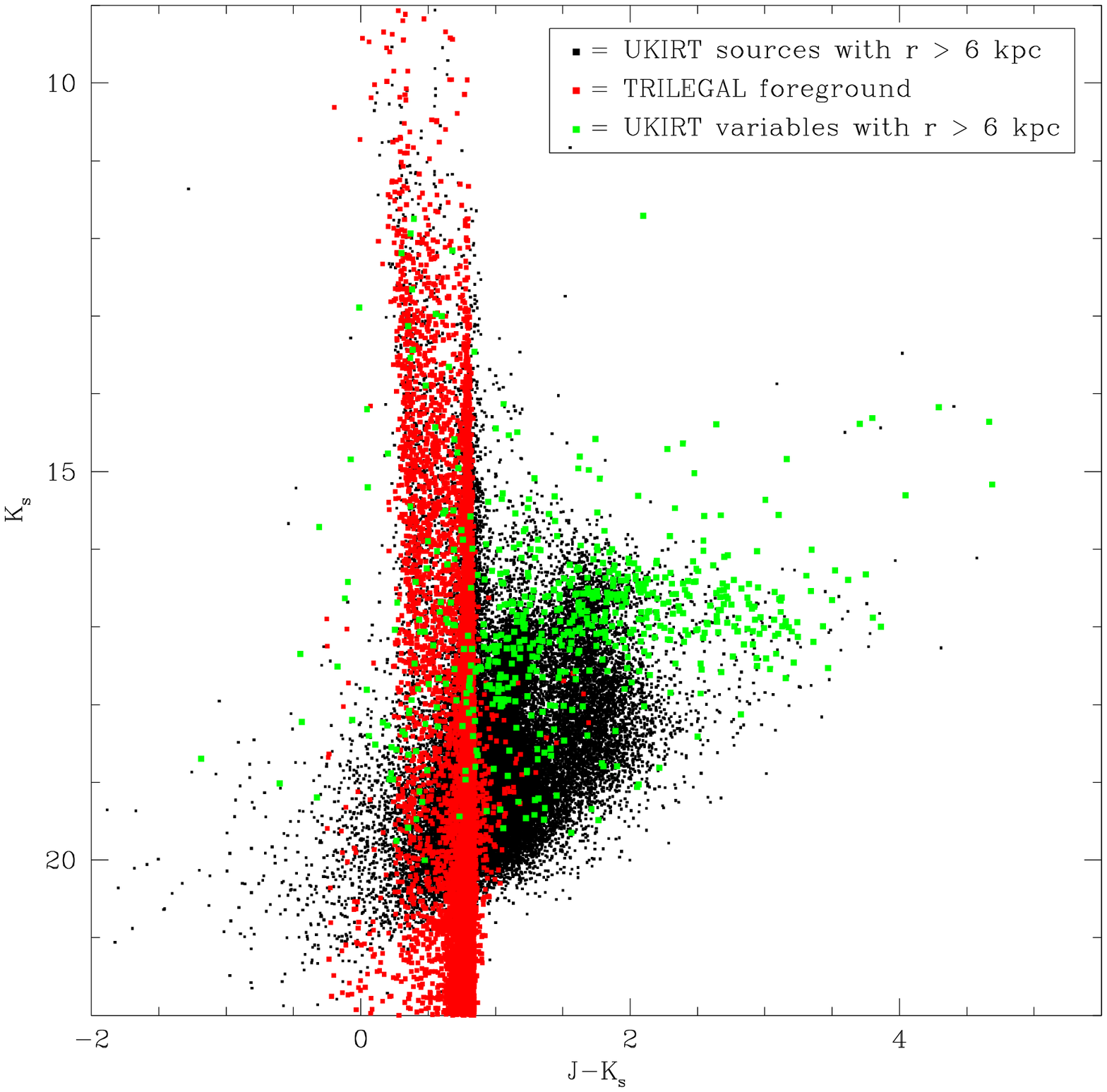,width=84mm}}
\caption[]{The UKIRT sources (black; variables in green) at $r>6$ kpc in the
galaxy plane. Overplotted in red is a TRILEGAL simulation for the same area of
0.68 square degrees.}
\end{figure}

% FIGURE 12
\begin{figure}
\epsfig{figure=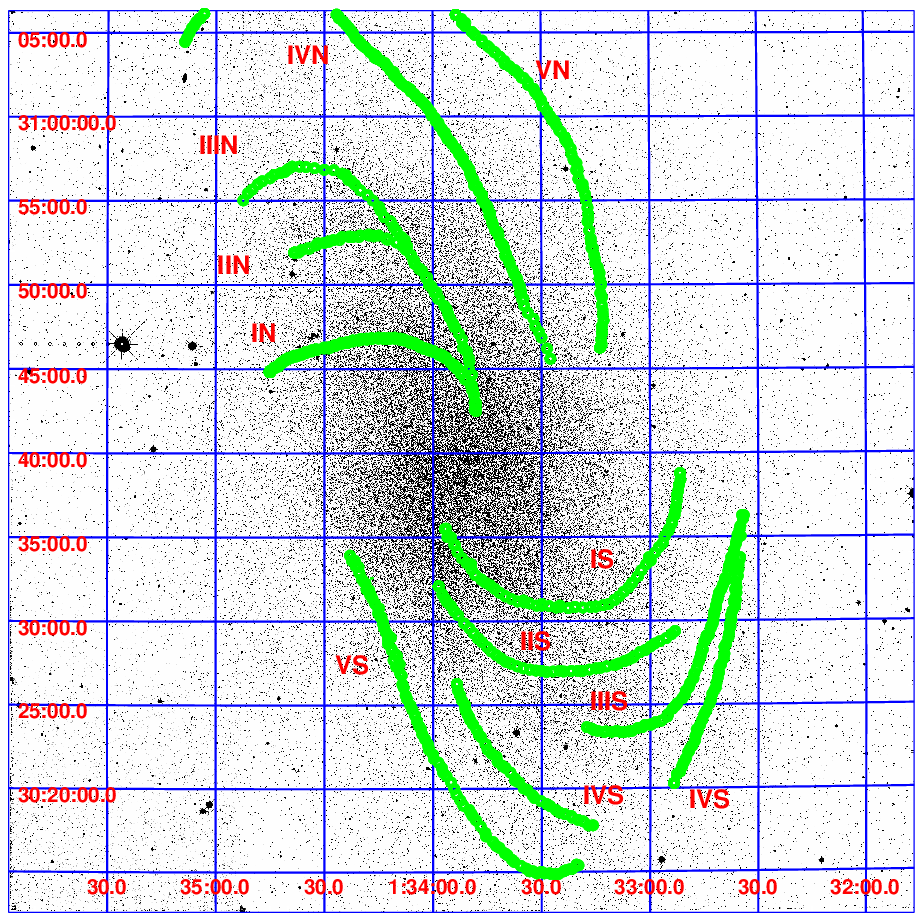,width=84mm}
\caption[]{WFCAM K-band mosaic of M\,33 with the system of five sets of spiral
arms marked on it.}
\centering
\end{figure}

% FIGURE 13 (REMOVED)
%\begin{figure}
%\centerline{\psfig{figure=fig13.ps,width=84mm}}
%\caption[]{Time lag $t$ of UKIRT variable stars with respect to arm II\,N,
%versus radial distance $r$. Solid green lines demarcate the adopted exclusion
%zone around the co-rotation radius (dotted line at $r=3.9$ kpc). We also
%investigated how the results would change for a slightly larger exclusion zone
%(outermost dotted lines) as well as for a limit to the acceptable values of
%$t$ ($-0.5<t<0.5$ Gyr) -- see Appendix.}
%\end{figure}

We divided our populations of LPVs into five age groups: (1) those formed
$t>10$ Gyr ago; (2) those formed between 5 Gyr $<t<10$ Gyr ago during the
first epoch of intense star formation; (3) those formed between 1 Gyr $<t<5$
Gyr ago, during the period of moderate star formation activity; (4) those
formed between 100 Myr $<t<1$ Gyr during the recent epoch of enhanced star
formation; and (5) massive (super-)AGB stars and RSGs formed $t<100$ Myr ago.
The Southern arms show a concentration near $t=0$ in all age bins. Arms I\,S
and II\,S show the strongest concentration of massive stars. The Northern arms
also show a concentration of stars, in all age bins, near $t=0$ but not
generally at $t=0$ itself. More surprisingly perhaps, there is no evidence
whatsoever for a lag associated with the density wave having passed through
the position where we see the stars dying now, or any asymmetry at all.

%-------------------------------------------------------------------------- 4.6
\subsection{Is there any structure?}

If spiral arms do not leave a lasting imprint on the distribution of young and
intermediate-age stellar populations, one could ask the question whether there
is any other structure which evolves with time? To this aim we construct the
distribution over nearest-neighbour separations (Fig.\ 14). The youngest stars
($t<100$ Myr) are indeed more agglomerated than the older groups of stars. The
100 Myr $<t<1$ Gyr population no longer shows any greater agglomeration than
still older populations, suggesting that the relaxation time for these
structures to dissolve is a few 100 Myr at most. To investigate this in more
detail, we divided the group of 100 Myr $<t<1$ Gyr stars into two subgroups:
those formed between 100 Myr $<t<300$ Myr (dotted line in Fig.\ 14), and those
formed between 300 Myr $<t<1$ Gyr (dashed line in Fig.\ 14). The younger of
these two groups still shows marginally greater agglomeration, implying a
relaxation time of $\sim200$ Myr (definitely slower than 100 Myr and faster
than 300 Myr).

%-------------------------------------------------------------------------- 4.7
\subsection{Correlation with ISM density}

Because stars form from dense gas, we would expect a correlation between the
ISM density and the density of young stars, but not old stars. Using the cold
dust density map constructed in Tabatabaei et al.\ (2014) from
{\it Herschel} Space Observatory and {\it Spitzer} Space Telescope data of
M\,33, we show in Fig.\ 15 histograms over ISM dust density, per each age
group defined previously. The different age populations exhibit very similar
distributions over density, probably because both stars and gas accumulate in
the deepest parts of the potential well (towards the centre, in the disc, and
in the spiral arms). The peak ISM dust density for all populations is around
$\log\rho({\rm g\ cm}^{-2})=-4.75$, which for a gas:dust mass ratio of
$\sim200$ (Braine et al.\ 2010) corresponds to ISM densities of $N\sim10^{21}$
cm$^{-2}$ -- typical of galaxy discs. The youngest stars ($t<100$ Myr),
however, are slightly biased towards higher ISM dust density, as was to be
expected.

%========================================================================== 5
\section{Discussion}
%-------------------------------------------------------------------------- 5.1
\subsection{Correction on pulsation duration}

In Paper III we realised that the duration of long period variability is
over-estimated by the theoretical Padova models, causing mass loss in excess
of the birth mass and an under-estimation of the SFR in our method. While we
determined an approximate correction factor $\sim10$ (cf.\ Rezaeikh et al.\
2014), this issue needs to be revisited in the light of the new data and
larger area of our survey. First, we examine how the SFRs we determined
compare with those determined using other methods.

% FIGURE 13
\begin{figure*}
\epsfig{figure=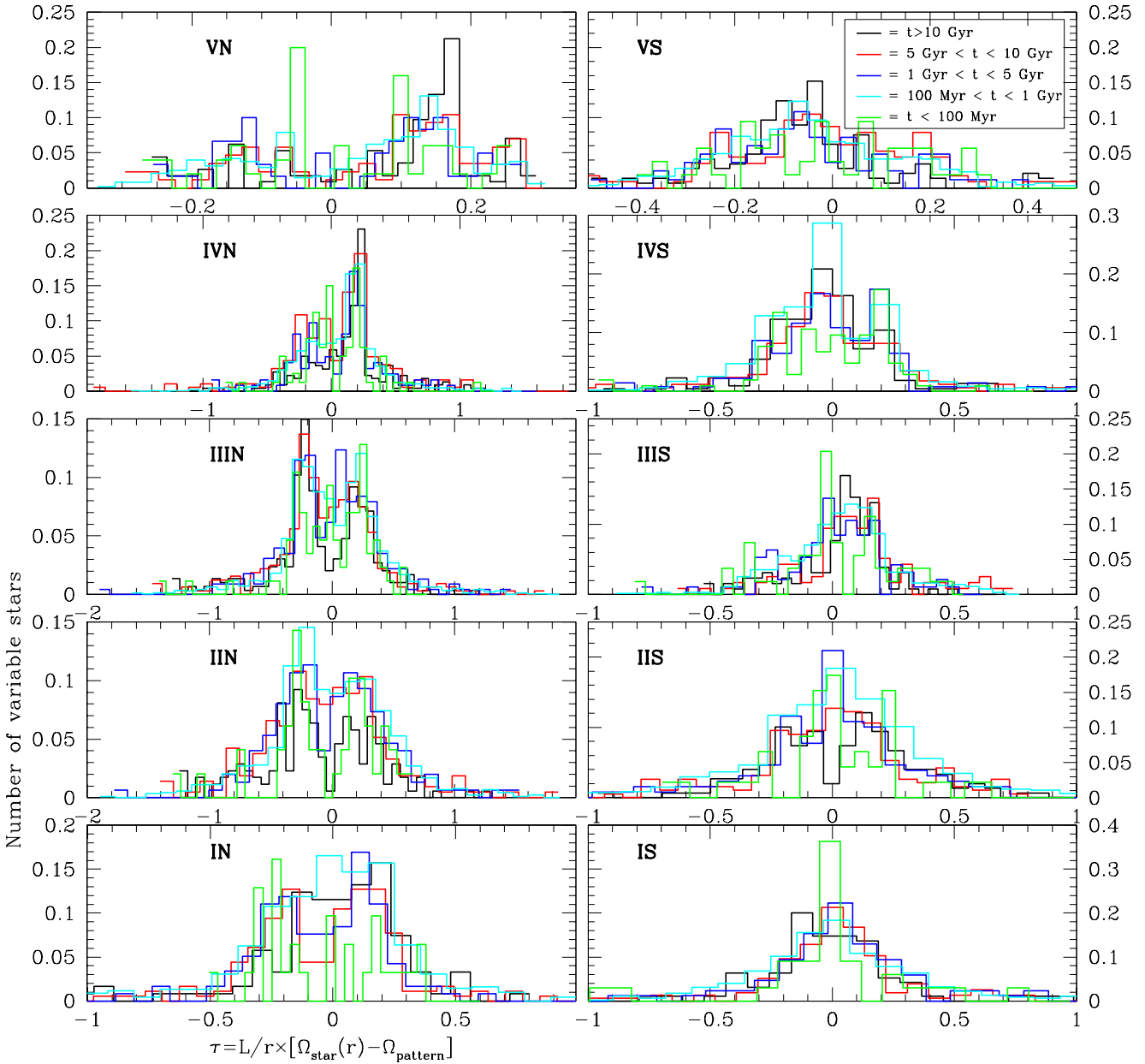,width=175mm}
\caption[]{Time lag (in Gyr) of variable stars formed at different times with
respect to five sets of spiral arms in the North and South of the disc of
M\,33. The calculations exclude the region around the co-rotation radius,
$r=3.9\pm0.3$ kpc.}
\centering
\end{figure*}

The current or recent SFR in M\,33 has been estimated from H$\alpha$ emission,
or the far-UV or far-IR luminosity by numerous groups (e.g., Hippelein et al.\
2003; Engargiola et al.\ 2003; Heyer et al.\ 2004; Gardan et al.\ 2007;
Boissier et al.\ 2007; Verley et al.\ 2009; Kang et al.\ 2012). Those SFRs
range mostly between 0.02--0.04 M$_\odot$ yr$^{-1}$ kpc$^{-2}$ for the central
regions of M\,33. Our new estimate of the current SFR in the central square
kpc, which is based on only a small number of RSGs, $\xi=0.007\pm0.003$
M$_\odot$ yr$^{-1}$ kpc$^{-2}$ (Fig.\ 4) is 3--6 times lower. Verley et al. \
(2009) estimated the SFR over the past 100 Myr across the disc of M\,33 to be
$\xi=0.45\pm0.10$ M$_\odot$ yr$^{-1}$. Our estimate for the entire disc of M\,33
(cf.\ Fig.\ 6, multiplied by the area covered) is $\xi=0.06\pm0.04$ M$_\odot$
yr$^{-1}$, i.e.\ 7--8 times lower. We thus conclude that our SRFs require
multiplication by a factor $\sim7$ to be in line with other, independent
estimates.

Is there any evidence for this correction factor to depend on mass and/or
metallicity? The SFRs for different epochs across the disc of M\,33 are shown
in Fig.\ 16, assuming the same correction factor of 7 applies to all. If this
is correct, then we see that for $t>4$ Gyr ago, the SFR reaches a maximum
value for $r<2$ kpc and further out it diminishes. At intermediate times,
$t\sim1$ Gyr, the radial profile of the SFR becomes much flatter, and in more
recent times the SFR peaks somewhere in the middle of the disc, around
$r\sim3$--5 kpc (i.e.\ not far from the co-rotation radius -- if there is such
a thing as a density wave). How does this compare to independent estimates?
Williams et al.\ (2009) determined the SFH based on {\it Hubble} Space
Telescope images of four fields located at $r=0.9$, 2.5, 4.3 and 6.1 kpc. The
innermost field was characterised by a first epoch of intense star formation
at $t\sim$10--14 Gyr ago, declining to $\xi=0.009$ M$_\odot$ yr$^{-1}$
kpc$^{-2}$ before increasing again to $\xi=0.045$ M$_\odot$ yr$^{-1}$ kpc$^{-2}$
at $t\sim$1--3 Gyr ago. The (corrected) SFR at that location derived from our
WFCAM data is $\xi\sim0.01$--0.02 M$_\odot$ yr$^{-1}$ kpc$^{-2}$ over the past
0.5--6 Gyr. At $t\sim10$ Gyr our SFR is lower, possibly due to incompleteness
at low masses. Around $r=2.5$ kpc the HST analysis suggested a much steadier
SFR without the initial ``burst'' of star formation, around $\xi=0.006$
M$_\odot$ yr$^{-1}$ kpc$^{-2}$. This compares favourably with our estimates at
$t=1$ and 5 Gyr, of $\xi\sim0.003$--0.18 M$_\odot$ yr$^{-1}$ kpc$^{-2}$ (Fig.\
16). Further out in the disc, Williams et al. (2009) suggest star formation
started at a low rate and gradually increased over time. This is very similar
to what we see (Figs.\ 7, 16 and A2). Around $r=4.3$ kpc, at $t=2$--8 Gyr ago
Williams et al.\ (2009) found $\xi\sim0.003$ M$_\odot$ yr$^{-1}$ kpc$^{-2}$,
indistinguishable from our result (Fig.\ 16); the same is true around $r=6.1$
kpc. Apart from the oldest stars in the central region, where a larger
correction factor might be needed, the consistency with the HST analysis
strongly affirms the use of a constant correction factor of 7 for all
populations across the disc of M\,33.

% FIGURE 14
\begin{figure}
\centerline{\psfig{figure=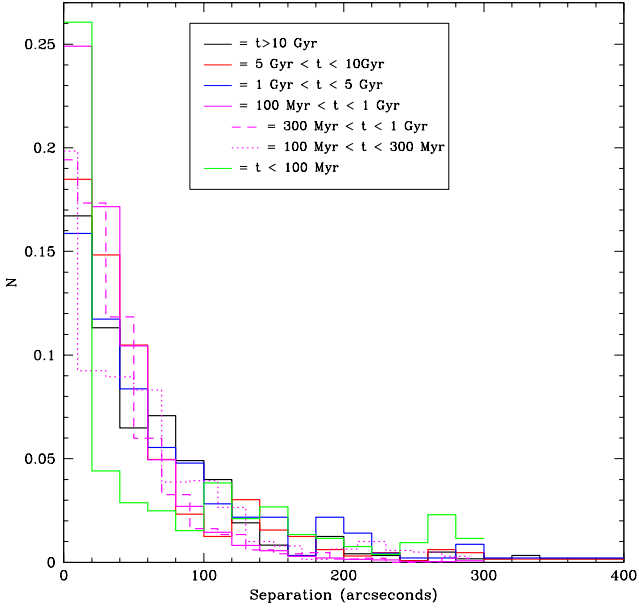,width=84mm}}
\caption[]{The nearest-neighbour separation distribution of UKIRT variable
stars formed at different times. Numbers of stars are normalised to total
numbers of stars in each age group.}
\end{figure}

% FIGURE 15
\begin{figure}
\centerline{\psfig{figure=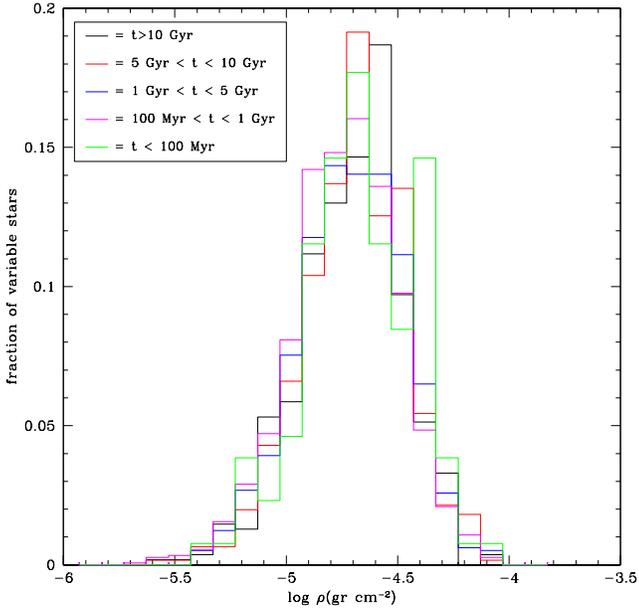,width=84mm}}
\caption[]{The ISM dust density distribution of UKIRT variable stars formed at
different times. Numbers of stars are normalised to total numbers of stars in
each age group.}
\end{figure}

% FIGURE 16
\begin{figure}
\centerline{\psfig{figure=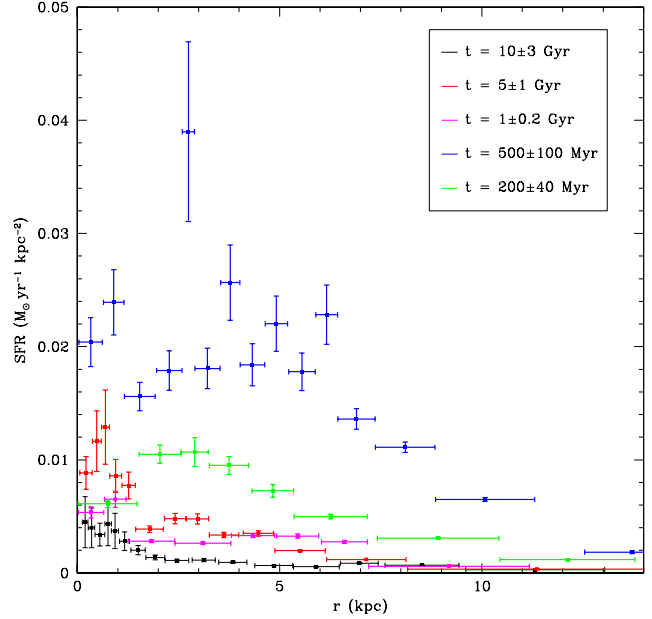,width=84mm}}
\caption[]{SFR across the disc of M\,33 at different times. A correction
factor of 7 has been applied to the SFRs.}
\end{figure}

In another study, Barker et al.\ (2007) observed three fields located at
projected radii of $\sim9$--13 kpc, SouthEast of M\,33's nucleus. They
reported a peak in star formation at $t\sim3$--5 Gyr ago ($\log t=9.4$--9.7)
for their fields A1 and A2 with a SFR of $\xi\sim1.2\times10^{-4}$ M$_\odot$
yr$^{-1}$ kpc$^{-2}$ and $\xi\sim2.8\times10^{-5}$ M$_\odot$ yr$^{-1}$
kpc$^{-2}$, respectively. Apart from that peak, the SFRs are typically 3.5,
1.8 and $0.8\times10^{-5}$ M$_\odot$ yr$^{-1}$ kpc$^{-2}$, for their fields
A1--3. At $r=9$ kpc, $t=3$--5 Gyr ago, we derive $\xi\sim1\times10^{-4}$
M$_\odot$ yr$^{-1}$ kpc$^{-2}$ (corrected; see Fig.\ A1); further out the SFR
drops to $\xi\sim2\times10^{-5}$ M$_\odot$ yr$^{-1}$ kpc$^{-2}$. These rates
compare very favourably with the results obtained by Barker et al.\ (2007),
vindicating the correction by a factor 7.

The question arises: to what extent is the correction factor due to survey
incompleteness? We can address this by looking at the central square kpc of
M\,33. The UIST survey (Paper I) found $N_{\rm U}=812$ variable stars; the
WFCAM survey (Paper IV) found $N_{\rm W}=667$. They had $N_{\rm UW}=192$ variable
stars in common. This already suggests that the UIST survey is no more
complete than $N_{\rm U}/(N_{\rm U}+N_{\rm W}-N_{\rm UW})\times100=63$\%, and
likewise the WFCAM survey is no more complete than 52\%. This is most likely
due to the limited sampling of the lightcurves, as most of the variable stars
that are sought have amplitudes that exceed the photometric uncertainties;
blending is unlikely to be a major factor as WFCAM recovers 95\% of the stars
(variable and not variable) detected with the sharper UIST camera. Thus, the
survey incompleteness of the UIST and WFCAM for the large-amplitude LPVs is
likely to be close to 60\% and 50\%, respectively, leaving another factor
3.5 that we attribute to the over-estimation by the models of the pulsation
duration. The latter correction cannot be much smaller than that -- and,
consequently, the survey incompleteness cannot be much worse than stated
above -- as the mass-loss catastrophy that first triggered the inquest into
the correction factor (Paper III) required a total correction of a factor 10.
(We expect that we need to seek to reduce the mass loss estimates by a factor
0.7, for the mass-loss catastrophy to be avoided when adopting a correction to
the number of variable stars of a factor 7.)

\subsection{Effect of changing the metallicity of older stars on SFH}

     As time passes, the metallicity of the ISM -- and hence that of new
     generations of stars -- changes as a result of nucleosynthesis and
     feedback from dying stars. Older stars are thus expected to have formed
     in more metal poor environments than younger stars, though this
     Age--Metallicity relation may vary spatially within a galaxy. Most of the
     chemical enrichment occurred $>5$ Gyr ago, when a significant fraction of
     the total stellar mass was put in place (note that chemical enrichment
     naturally slows down dramatically once the metallicity approaches solar
     values). While this means that the assumed metallicity in section 4 is
     quite reliable for young and intermediate-age populations, the oldest
     populations ($>6$ Gyr) might be deficient in metals.

     To examine the metallicity effect on the early SFH, we consider the
     following metallicities: $Z=0.008$, 0.004, 0.0024, 0.0019, 0.0015 and
     0.0012. The parameterisations of the Mass--Luminosity relation,
     Mass--Pulsation relation and Mass-Age relation for these different
     metallicities are provided in the Appendix (see Golshan et al.\ 2016 for
     details of the fitting). While the incorporation in our method of an
     Age--Metallicity relation is non-trivial and beyond the scope of the
     present work, the resulting SFHs for different -- but uniform -- values
     of the metallicity are shown in Fig.\ 17. Interpretation of this diagram
     must take notice of what is known about the metallicities of populations
     of different ages within M\,33.

     Barker et al.\ (2007) used optical CMDs to determine the SFH and
     Age--Metallicity relation for three fields at deprojected radii of 9--13
     kpc -- we consider this the most extreme scenario for any deviation from
     our assumed metallicity. SFRs peaked $\sim4$ Gyr ago in the innermost two
     fields but $>6$ Gyr in the outermost field. Their metallicities are
     $Z\sim0.004$--0.006 for most of the past $\sim6$ Gyr ($\log t<9.8$), but
     they were $Z\sim0.0024$--0.003 around $t\sim10$ Gyr.

     For $t<5$ Gyr ago ($\log t<9.7$) we see (in Fig.\ 17) that changing the
     adopted metallicity from $Z=0.008$ to $Z=0.004$ has a negligible effect
     on the derived SFR. Such change also does not affect the SFRs at the
     oldest ages ($t>10$ Gyr), but it does increase the SFRs at $t\sim6$ Gyr
     ago by up to a factor three. The change is not very much larger for a
     metallicity as low as $Z=0.0024$ or even as low as $Z=0.0019$. Certainly,
     while the derived SFRs in the period $6<t<10$ Gyr ago ($9.8<\log t<10$)
     would increase by an order of magnitude if the metallicity is as low as
     $Z=0.0012$--0.0015, such low metallicities are not appropriate even for
     the relatively metal-poor outskirts of the M\,33 disc.

     In conclusion, the results we have derived by adopting $Z=0.008$ across
     the disc of M\,33 are robust against reasonable deviations from this
     value for the metallicity, with a possible modest underestimate of the
     star formation around $t\sim6$--8 Gyr ago.

% FIGURE 17
\begin{figure}
\centerline{\psfig{figure=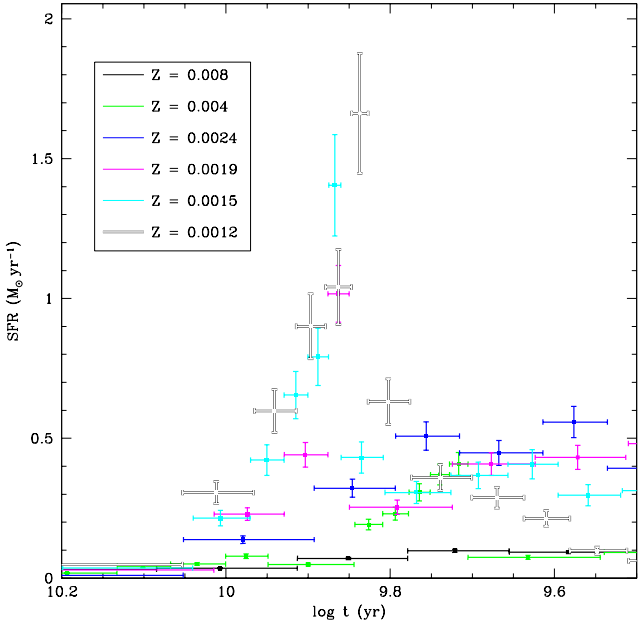,width=84mm}}
\caption[]{
 SFH compared for six choices of metallicity.
}
\end{figure}

%-------------------------------------------------------------------------- 5.2
\subsection{The SFH in M\,33}

In the following, all SFRs are corrected by a factor 7.

   The star formation rate has varied between $\sim0.010\pm0.001$ 
   ($\sim0.012\pm0.007$) and 0.060$\pm0.005$ (0.052$\pm0.009$) M$_\odot$ 
   yr$^{-1}$ kpc$^{-2}$ statistically (systematically) in the central square
   kiloparsec of M\,33. The total star formation rate in M\,33 within a galactocentric
   radius of 14 kpc has varied between $\sim0.110\pm0.005$ ($\sim0.174\pm0.060$)
   and $\sim0.560\pm0.028$ ($\sim0.503\pm0.100$) M$_\odot$ yr$^{-1}$ statistically
   (systematically). The statistical estimation of SFRs with associated star number 
   statistics (represented by the vertical errorbars) are shown for the central region and 
   the disc in Fig.\ 4 and Fig.\ 6, respectively. The systematic estimates of 
   SFRs reflect the variation in SFRs arising from the initial assumptions such as 
   the model we choose, the distance modulus we adopt and the metallicity we adopt. 
   Each of these factors has been investigated carefully  in this paper and/or in 
   paper II (see Fig.\ 10 in paper II for the effect of distance modulus, Fig.\ 11 
   in paper II and Fig.\ 17 in this paper for the metallicity effect and Fig.\ A4 for
   the model effect). 

We find that the total mass of stars formed in the central square kpc of M\,33
is $M_\star=2.1\times10^8$ M$_\odot$, of which $1.53\times10^8$ M$_\odot$ (73\%)
was formed $t>4$ Gyr ago ($\log t>9.6$), but only $1.36\times10^7$ M$_\odot$
(6.5\%) was formed in the most recent few hundred Myr. These values agree with
our previous results based on the UIST data (Paper II). In the disc of M\,33,
we find a total mass of stars $M_\star=1.7\times10^9$ M$_\odot$, of which
$1.21\times10^9$ M$_\odot$ (71\%) was formed $t>3$ Gyr ago ($\log t>9.5$), but
only $2.26\times10^8$ (13\%) was formed in the most recent few hundred Myr.
The total mass in stars is at the low end of the range given by Corbelli et
al.\ (2003); we probably missed a significant number of the oldest,
lowest-mass stars that do not develop large amplitude pulsation.

Our results suggest that, currently, star formation is more active in the disc
than in the central square kpc, and vice versa in the early assembly of M\,33
-- an inside--out formation scenario. Of course, this does not take into
account the effect of stellar migration, which may be important for the older
populations and most easily discerned in the outskirts where no new stars
form (cf.\ El-Badry et al.\ 2015; Miranda et al.\ 2015).

The recent enhancement in the SFR confirms the result of Minniti et al.\
(1993), who found that the central regions in the disc of M\,33 experienced
enhanced star formation activity over the past 1 Gyr. This may have been
caused by interaction with the much more massive Andromeda spiral galaxy,
M\,31 (Richardson et al.\ 2011). Evidence for interaction comes from the
warped shape of the disc (Sandage \& Humphreys 1980), and could also have
caused the accretion of gas into the central regions necessary to sustain the
star formation at the rates observed (Paper III).

Hagen et al.\ (2015) determined the recent SFH on the basis of UV photometry.
The SFR has varied between 0.01 and 1.4 M$_\odot$ yr$^{-1}$ over the past 400
Myr, with the higher values associated with peaks in the SFR around 4 and 10
Myr ago. This is partially due to the better time resolution in recent times,
with past such peaks having been smeared out in time to much lower averages.
Our estimate of 0.42 M$_\odot$ yr$^{-1}$ over the past 100 Myr is in good
agreement with their results.

Kang et al.\ (2012) constructed a parametrized model to describe the main
characteristics of the M\,33 SFH, assuming a moderate outflow and inside--out
formation scenario. The latter agrees with the negative age gradient in the
inner disc measured by Williams et al.\ (2009), as well as our $\xi(t,r)$.

Beyond a certain ``break'' radius, the older populations regain prominence
(Barker et al. \ 2007). We find that in the central regions the peak in SFR is
similar between the old and recent star formation, but this ratio gradually
declines as one moves out through the disc of M\,33, reaching a minimum around
$r\sim8$ kpc where the ancient SFR is only $\sim10$\% of that in recent times.
At $r>8$ kpc the stellar population becomes older again. This was also seen
for 17 out of 44 face-on galaxies by Ruiz-Lara et al.\ (2015); they propose
that the disc of these systems were formed in their entirety at early times,
but that more recent star formation has been quenched from inside--out.

While younger stars ($t<100$ Myr) are more prevalent in the spiral arms of
M\,33, all stellar populations show a similar concentration towards these
arms, with a conspicuous lack of any asymmetry. This means that while stars
migrate out of the arms they do not systematically lag behind them. This, in
turn, means that the spiral arms do not reflect a density wave potential
which has a pattern speed that is notably different from that of the stellar
orbital velocities. We conclude that the spiral arms are transient features,
in accordance with numerical simulations (e.g., Grand et al.\ 2012). Junichi
et al.\ (2015) also performed three-dimensional $N$-body/hydrodynamic
simulations and found that the spiral pattern in barred galaxies could change
on time scales of 100 Myr, washing out any relation between older stars and
spiral arms. This means that grand-design spiral arms in barred galaxies are
not stationary, but transient. Our results suggest this is also the case in
lower-mass, non-barred spiral galaxies; stars are accumulated temporarily
within spiral arms, but over time spans of 100 Myr or more the star formation
takes place throughout the disc so stars quickly lose memory of where they
were formed.

%========================================================================== 6
\section{Conclusions}

The photometric catalogue of near-IR variable stars of Paper IV was used to
reconstruct the SFH -- following the method introduced in Paper II -- across
the galactic disc of M\,33 and with respect to its spiral arm pattern. We
summarise the main results as follows:

\begin{itemize}
\item{Most stars in M\,33, $\approx71$\% were formed $\sim5$ Gyr ago, i.e.\
redshift $\sim1$.}
\item{The star formation history of M\,33 has peaked again more recently, and
we detect one such peak $\sim250$ Myr ago, lasting $\sim200$ Myr and forming
$\sim13$\% of all stars in M\,33.}
\item{From the centre outwards through the disc of M\,33, the early epoch of
star formation becomes progressively less pronounced, indicating a youtfull
disc consistent with an inside--out formation scenario.}
\item{Old stars become more prominent again in the outskirts of M\,33, at
$>8$ kpc from the centre, possibly indicating a reduced star formation
efficiency in the low density gas in combination with a dynamically relaxed,
more extended old stellar population.}
\item{We find that stars of all ages trace some of the spiral arms, but only
stars younger than $\sim100$ Myr show a greater concentration in the arms;
likewise only stars younger than $\sim100$ Myr show a greater affinity with
high density gas. This indicates that dynamical mixing operates on timescales
$<100$ Myr.}
\item{We determine a survey incompleteness of $\sim50$\%, and suggest the
stellar evolution models we use may have overestimated the lifetimes of the
large-amplitude LPV phase by a factor 3 to 4.}
\end{itemize}

%========================================================================== A
\appendix 
%-------------------------------------------------------------------------- A1
\section{Supplementary material}

Here, we show additional diagrams referred to in the text, providing
alternative or additional information.

In Fig.\ A1, we show the SFH assuming all stars are oxygen-rich, i.e.\ we
neglect the fact that stars in a certain mass range will be carbon stars with
somewhat different photometric properties. The SFH does not deviate from that
in Fig.\ 6 by a great deal, and our conclusions are therefore not sensitive to
the exact manner in which we assign the chemical class to the variable stars.

% FIGURE A1
\begin{figure}
\centerline{\psfig{figure=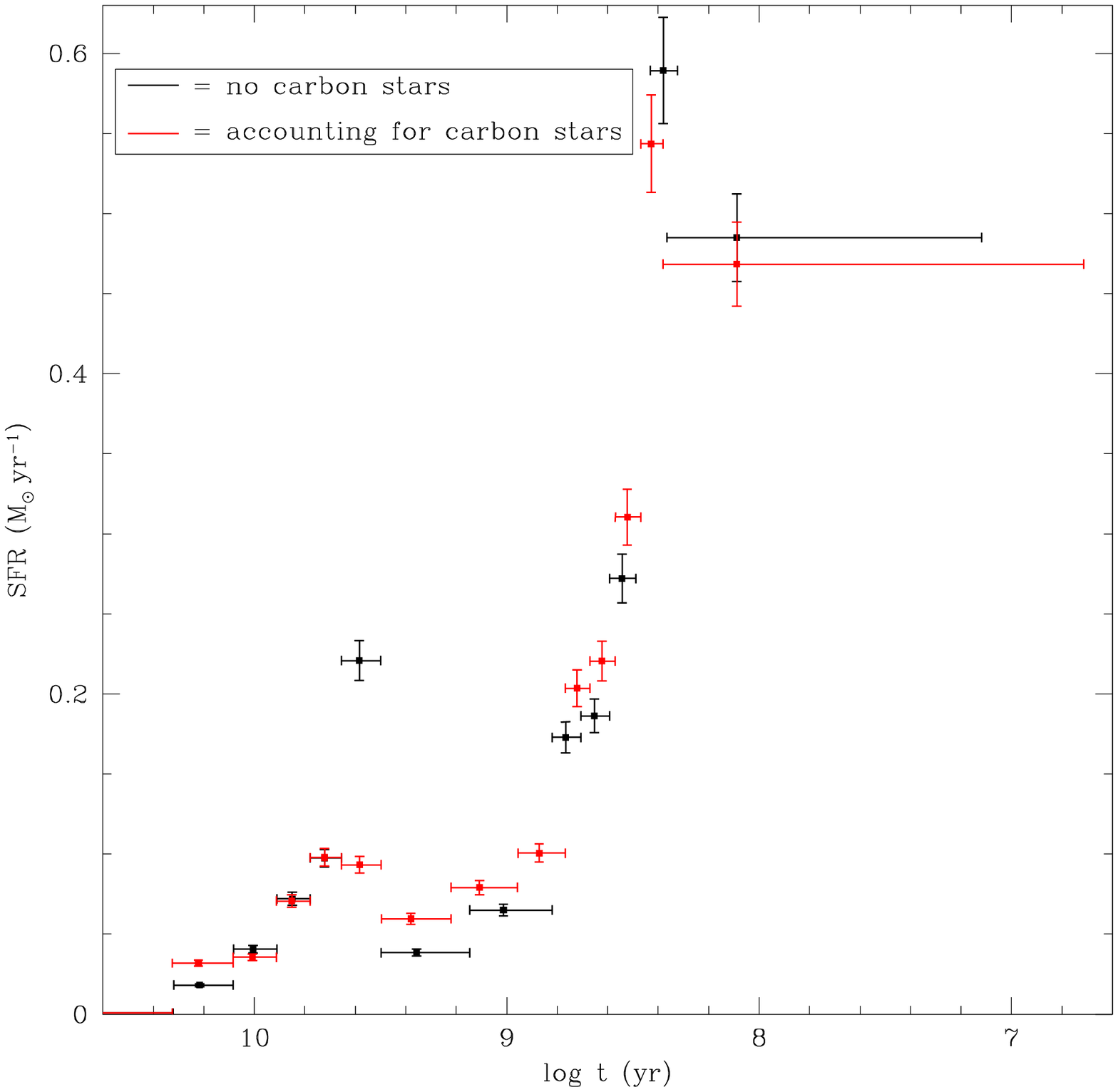,width=84mm}}
\caption[]{
     The SFH across the entire disc of M\,33, assuming all stars are
     oxygen-rich (black) compared to the SFH we derived assuming
     intermediate-mass stars will be carbon stars (red); a correction factor
     of 7 has been applied.
}
\end{figure}

In Fig.\ A2, we show how the SFH varies between regions at increasing distance
to the centre of M\,33. This contains similar information to that in Fig.\ 7,
but in a way which facilitates reading off the precise values for the SFR.

% FIGURE A2
\begin{figure}
\centerline{\psfig{figure=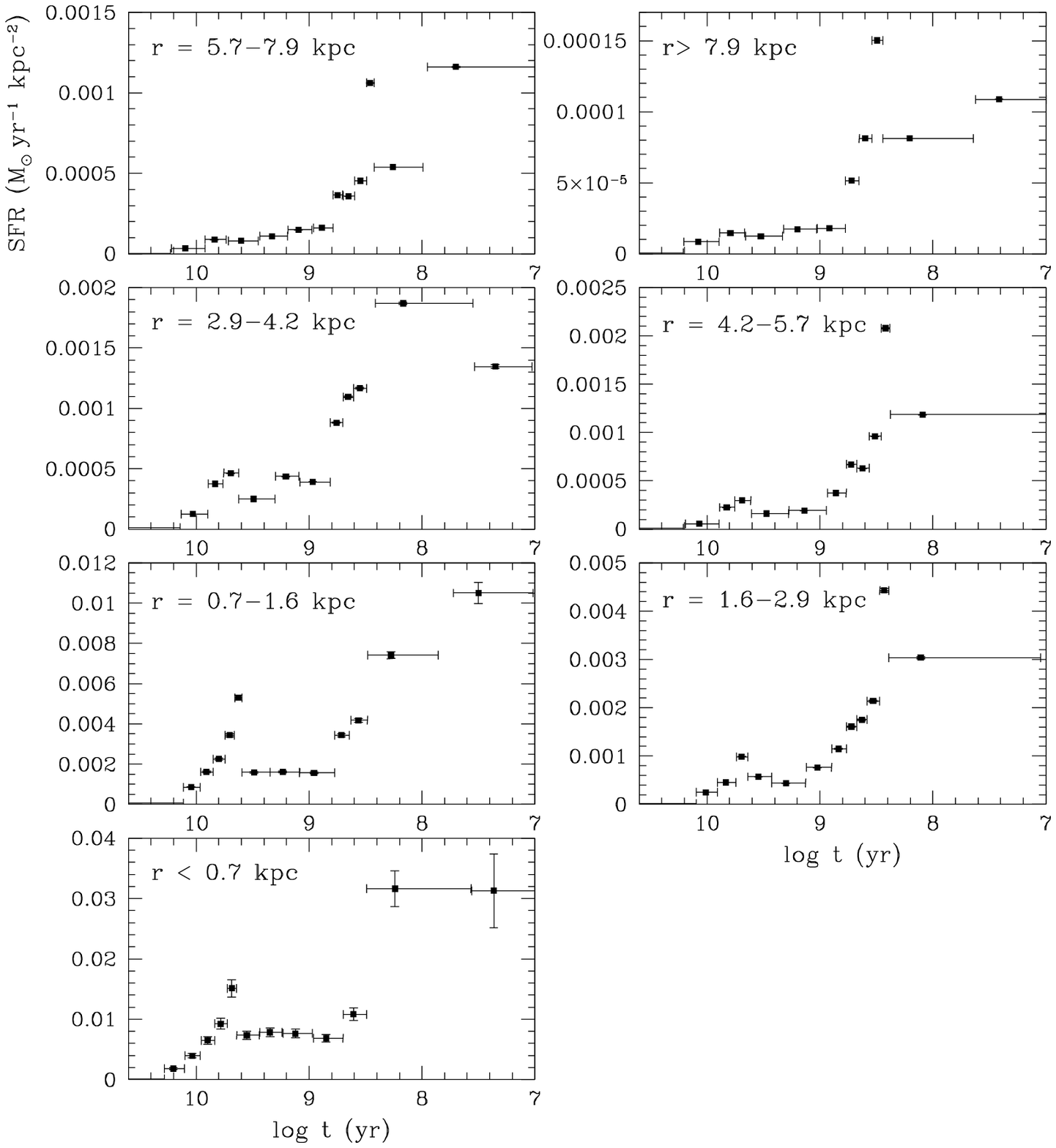,width=84mm}}
\caption[]{SFH of M\,33 in different radius bins, in the galaxy plane. A
correction factor of 7 has been applied.}
\end{figure}

In Fig.\ A3, we show the results of the investigation of the association of
stars of different ages with the spiral arms, employing a different (from that
in the main body of the paper) exclusion criterion to avoid the singularities
arising at the co-rotation radius. The conclusions are unaltered.

% FIGURE A3
\begin{figure}
\centerline{\psfig{figure=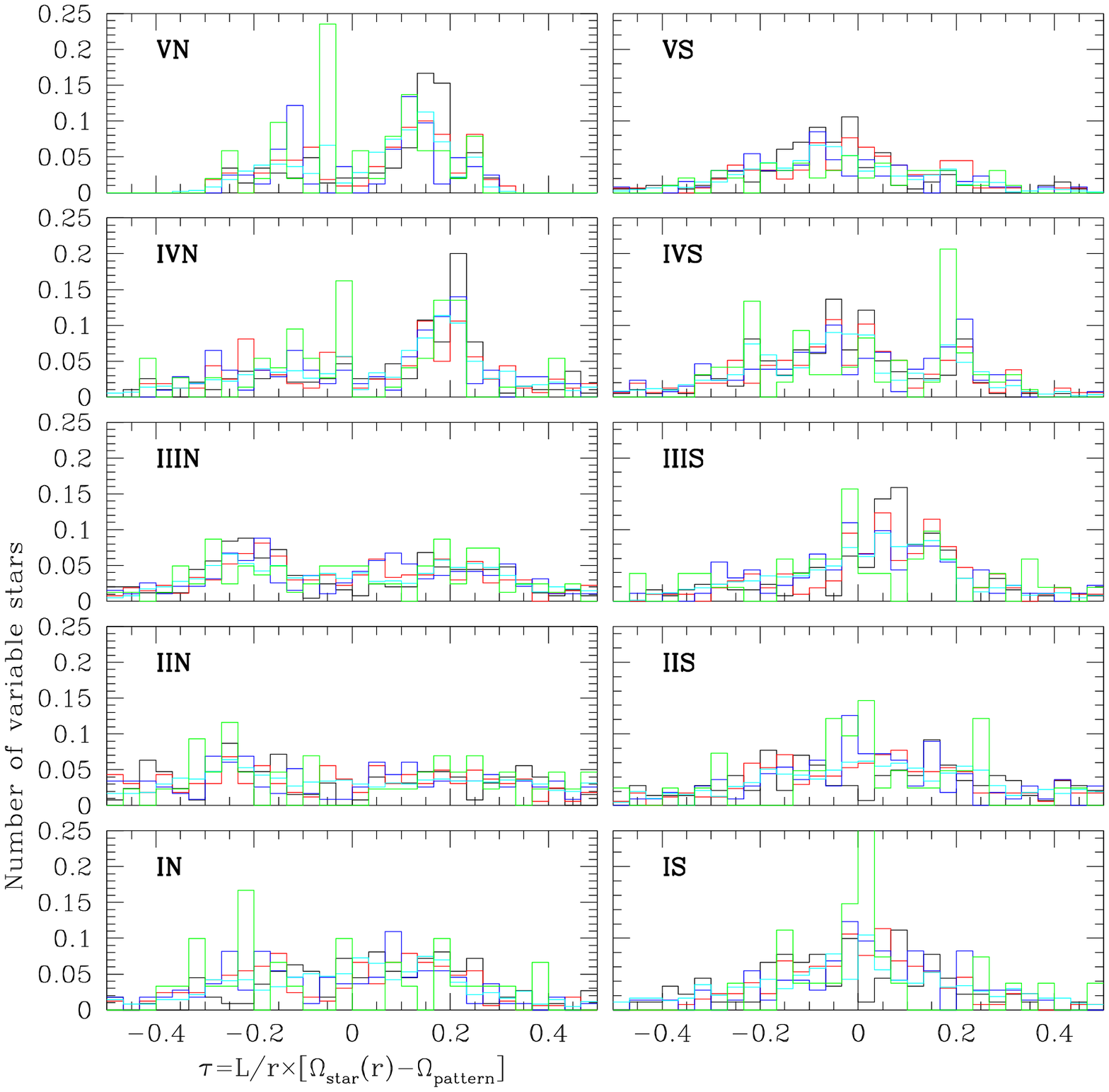,width=84mm}}
\caption[]{As figure 14, but now the calculations have been limited to time
lags of $-0.5<t<0.5$.
% (red lines in Fig.\ 13)
}
\end{figure}

     In Fig.\ A4, we compare the Mass--Luminosity relation (in the K-band) for
     $Z=0.008$ derived from Padova isochrones with BaSTI isochrones.

% FIGURE A4
\begin{figure}
\centerline{\psfig{figure=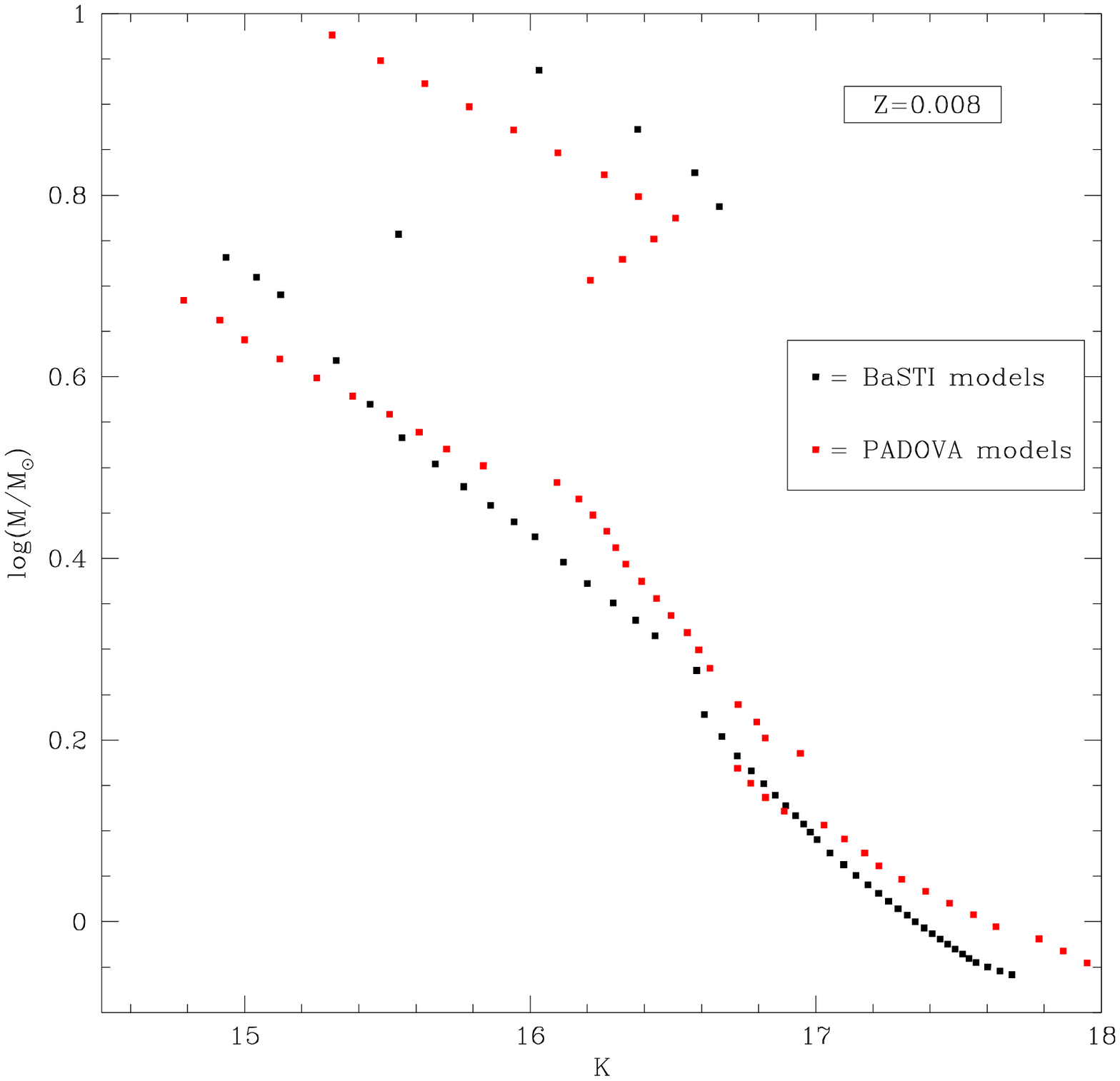,width=84mm}}
\caption[]{
     Mass--Luminosity relation (in the K-band) for $Z=0.008$ and a distance
     modulus of $\mu=24.9$ mag derived from the BaSTI isochrones (in black)
     and Padova isochrones (in red).
}
\end{figure}

     In Fig.\ A5, we compare the Mass--Age relation for $Z=0.008$ derived from
     Padova isochrones with BaSTI isochrones.

% FIGURE A5
\begin{figure}
\centerline{\psfig{figure=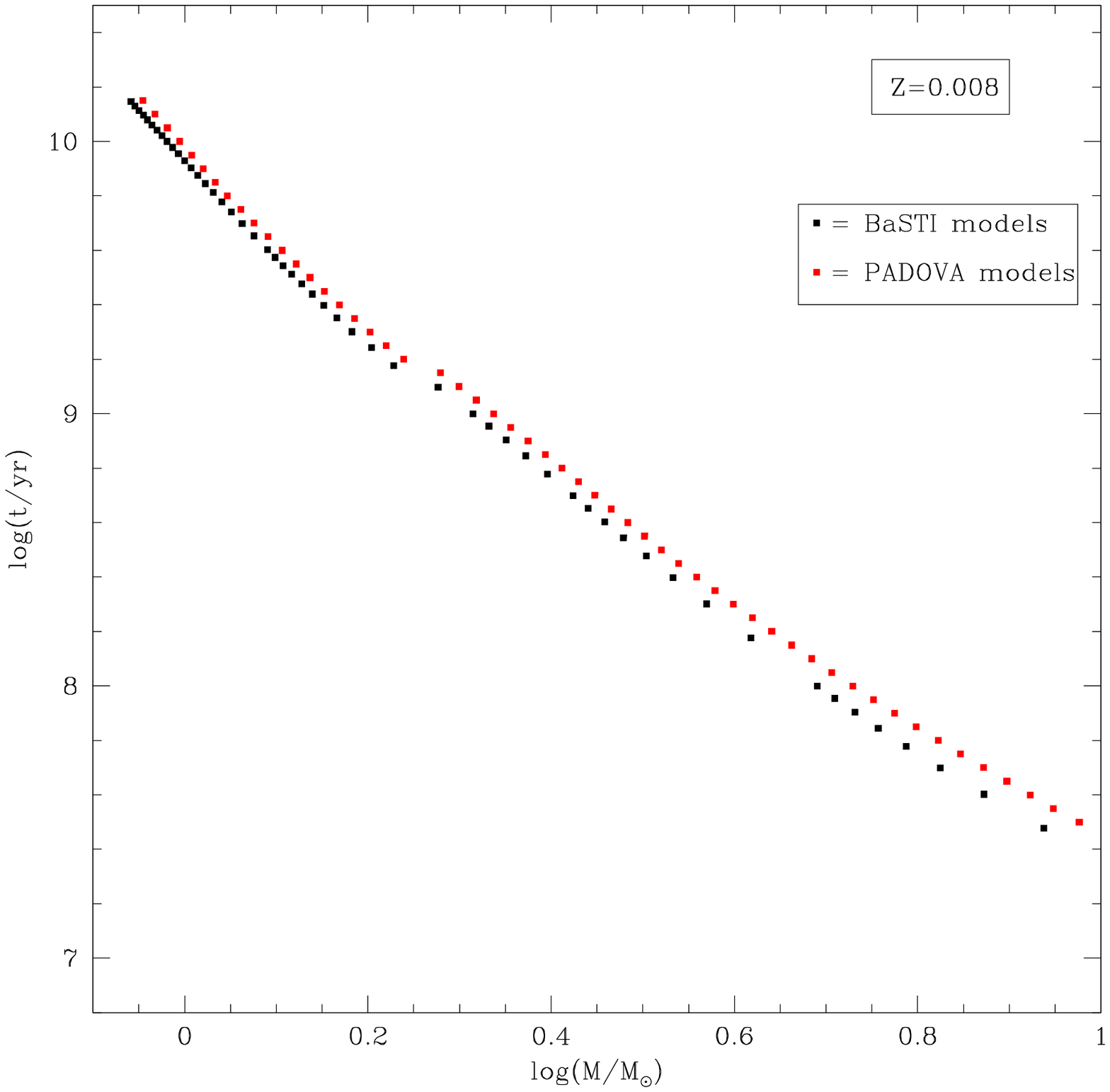,width=84mm}}
\caption[]{
    (Birth) Mass--Age relation of AGB stars for $Z=0.008$ derived from the
     BaSTI isochrones (in black) and Padova isochrones (in red).
}
\end{figure}

     In Tables A1, A2 and A3, we present parameterisations of the
     Mass--Luminosity relation, Mass--Age relation and Mass--Pulsation
     relation derived from the theoretical models of Marigo et al.\ (2008) and
     used in our derivation of the SFH in Sections 4 and 5.2.

% TABLE A1
\begin{table}
\caption[]{Relation between birth mass and $K$-band magnitude, $\log M=aK+b$,
for a distance modulus of $\mu=0$ mag.}
\begin{tabular}{ccr}
\hline\hline
$a$              & $b$              & validity range             \\ 
\hline
\multicolumn{3}{c}{$Z=0.0012$} \\
\hline
$-0.447\pm0.086$ & $-3.502\pm0.985$ &         $K\leq    -11.051$ \\
$-0.897\pm0.084$ & $-8.476\pm0.904$ & $-11.051<K\leq    -10.238$ \\
$-0.233\pm0.075$ & $-1.683\pm0.737$ & $-10.238<K\leq\ \, -9.426$ \\
$-0.085\pm0.071$ & $-0.286\pm0.643$ &  $-9.426<K\leq\ \, -8.614$ \\
$-0.231\pm0.062$ & $-1.542\pm0.503$ &  $-8.614<K\leq\ \, -7.802$ \\
$-0.315\pm0.053$ & $-2.198\pm0.392$ &  $-7.802<K\leq\ \, -6.989$ \\
$-0.080\pm0.063$ & $-0.553\pm0.419$ &         $K  > \ \, -6.989$ \\
\hline
\multicolumn{3}{c}{$Z=0.0015$} \\
\hline
$-0.447\pm0.071$ & $-3.477\pm0.808$ &         $K\leq    -11.006$ \\
$-0.918\pm0.066$ & $-8.667\pm0.698$ & $-11.006<K\leq    -10.204$ \\
$-0.232\pm0.050$ & $-1.663\pm0.486$ & $-10.204<K\leq\ \, -9.401$ \\
$-0.094\pm0.047$ & $-0.362\pm0.426$ &  $-9.401<K\leq\ \, -8.599$ \\
$-0.219\pm0.041$ & $-1.436\pm0.329$ &  $-8.599<K\leq\ \, -7.796$ \\
$-0.333\pm0.036$ & $-2.330\pm0.272$ &  $-7.796<K\leq\ \, -6.994$ \\
$-0.062\pm0.046$ & $-0.432\pm0.301$ &         $K  > \ \, -6.994$ \\
\hline
\multicolumn{3}{c}{$Z=0.0019$} \\
\hline
$-0.508\pm0.052$ & $-4.152\pm0.676$ &         $K\leq    -10.998$ \\
$-0.893\pm0.048$ & $-8.380\pm0.649$ & $-10.998<K\leq    -10.175$ \\
$-0.234\pm0.048$ & $-1.677\pm0.688$ & $-10.175<K\leq\ \, -9.383$ \\
$-0.095\pm0.044$ & $-0.362\pm0.673$ &  $-9.383<K\leq\ \, -8.590$ \\
$-0.238\pm0.038$ & $-1.597\pm0.606$ &  $-8.590<K\leq\ \, -7.798$ \\
$-0.317\pm0.032$ & $-2.219\pm0.534$ &  $-7.798<K\leq\ \, -7.000$ \\
$-0.076\pm0.042$ & $-0.524\pm0.471$ &         $K  > \ \, -7.000$ \\
\hline 
\multicolumn{3}{c}{$Z=0.0024$} \\
\hline
$-0.475\pm0.070$ & $-3.777\pm0.787$ &         $K\leq    -10.926$ \\
$-0.911\pm0.061$ & $-8.538\pm0.645$ & $-10.926<K\leq    -10.143$ \\
$-0.225\pm0.063$ & $-1.585\pm0.612$ & $-10.143<K\leq\ \, -9.359$ \\
$-0.098\pm0.060$ & $-0.393\pm0.536$ &  $-9.359<K\leq\ \, -8.576$ \\
$-0.253\pm0.051$ & $-1.722\pm0.412$ &  $-8.576<K\leq\ \, -7.792$ \\
$-0.256\pm0.042$ & $-1.743\pm0.308$ &  $-7.792<K\leq\ \, -7.009$ \\
$-0.147\pm0.057$ & $-0.985\pm0.382$ &         $K  > \ \, -7.009$ \\
\hline
\multicolumn{3}{c}{$Z=0.004$} \\
\hline
$-0.544\pm0.087$ & $-4.621\pm0.971$ &         $K\leq    -10.819$ \\
$-0.803\pm0.074$ & $-7.423\pm0.771$ & $-10.819<K\leq    -10.083$ \\
$-0.215\pm0.073$ & $-1.492\pm0.707$ & $-10.083<K\leq\ \, -9.347$ \\
$-0.076\pm0.070$ & $-0.194\pm0.634$ &  $-9.347<K\leq\ \, -8.610$ \\
$-0.321\pm0.061$ & $-2.304\pm0.501$ &  $-8.610<K\leq\ \, -7.874$ \\
$-0.213\pm0.052$ & $-1.452\pm0.396$ &  $-7.874<K\leq\ \, -7.138$ \\
$-0.142\pm0.071$ & $-0.949\pm0.489$ &         $K  > \ \, -7.138$ \\
\hline
\multicolumn{3}{c}{$Z=0.008$} \\
\hline
$-0.840$         &  \llap{$-8$}.076 &         $K\leq    -11.168$ \\
$-0.589$         &  \llap{$-5$}.275 & $-11.168<K\leq    -10.113$ \\
$-0.188\pm0.028$ & $-1.216\pm0.043$ & $-10.113<K\leq\ \, -9.600$ \\
$-0.142\pm0.027$ & $-0.778\pm0.041$ & $ -9.600<K\leq\ \, -9.087$ \\
$-0.188\pm0.022$ & $-1.194\pm0.036$ & $ -9.087<K\leq\ \, -8.573$ \\
$-0.501\pm0.016$ & $-4.291\pm0.024$ & $ -8.573<K\leq\ \, -8.060$ \\
$-0.248\pm0.018$ & $-1.840\pm0.026$ & $ -8.060<K\leq\ \, -7.547$ \\
$-0.128\pm0.025$ & $-0.930\pm0.035$ &         $K  > \ \, -7.547$ \\
\hline
\end{tabular}
\end{table}

%TABLE A1 continued
\begin{table}
\contcaption{}
\begin{tabular}{ccr}
\hline\hline
\multicolumn{3}{c}{$Z=0.015$} \\
\hline
$-2.198$         & \llap{$-23$}.425 &         $K\leq    -11.267$ \\
$-0.535$         &  \llap{$-4$}.696 & $-11.267<K\leq    -10.046$ \\
$-0.213\pm0.027$ & $-1.459\pm0.040$ & $-10.046<K\leq\ \, -9.553$ \\
$-0.088\pm0.018$ & $-0.267\pm0.023$ &  $-9.553<K\leq\ \, -9.060$ \\
$-0.222\pm0.021$ & $-1.472\pm0.024$ &  $-9.060<K\leq\ \, -8.568$ \\
$-0.582\pm0.012$ & $-4.554\pm0.014$ &  $-8.568<K\leq\ \, -8.075$ \\
$-0.147\pm0.013$ & $-1.040\pm0.016$ &  $-8.075<K\leq\ \, -7.582$ \\
$-0.173\pm0.022$ & $-1.247\pm0.032$ &         $K  > \ \, -7.582$ \\
\hline
\end{tabular}
\end{table}

% TABLE A2
\begin{table}
\caption[]{Relation between age and birth mass, $\log t=a\log M+b$.}
\begin{tabular}{ccr}
\hline \hline
$a$              & $b$             & validity range           \\
\hline
\multicolumn{3}{c}{$Z=0.0012$} \\
\hline
$-3.109\pm0.033$ & $9.842\pm0.003$ &       $\log{M}\leq0.241$ \\
$-2.433\pm0.032$ & $9.678\pm0.013$ & $0.241<\log{M}\leq0.555$ \\
$-2.043\pm0.034$ & $9.462\pm0.024$ & $0.555<\log{M}\leq0.868$ \\
$-1.618\pm0.039$ & $9.093\pm0.040$ & $0.868<\log{M}\leq1.181$ \\
$-1.074\pm0.046$ & $8.451\pm0.061$ & $1.181<\log{M}\leq1.495$ \\
$-0.833\pm0.060$ & $8.090\pm0.097$ &       $\log{M}>1.495$    \\
\hline
\multicolumn{3}{c}{$Z=0.0015$} \\
\hline
$-3.216\pm0.030$ & $9.860\pm0.002$ &       $\log{M}\leq0.198$ \\
$-2.511\pm0.028$ & $9.720\pm0.009$ & $0.198<\log{M}\leq0.465$ \\
$-2.238\pm0.030$ & $9.593\pm0.018$ & $0.465<\log{M}\leq0.732$ \\
$-1.850\pm0.033$ & $9.310\pm0.028$ & $0.732<\log{M}\leq0.998$ \\
$-1.413\pm0.038$ & $8.873\pm0.042$ & $0.998<\log{M}\leq1.265$ \\
$-1.063\pm0.043$ & $8.430\pm0.060$ & $1.265<\log{M}\leq1.531$ \\
$-0.801\pm0.054$ & $8.030\pm0.088$ &       $\log{M}>1.531$    \\
\hline
\multicolumn{3}{c}{$Z=0.0019$} \\
\hline
$-3.221\pm0.030$ & $9.874\pm0.002$ &       $\log{M}\leq0.200$ \\
$-2.511\pm0.027$ & $9.732\pm0.009$ & $0.200<\log{M}\leq0.465$ \\
$-2.278\pm0.029$ & $9.624\pm0.018$ & $0.465<\log{M}\leq0.729$ \\
$-1.856\pm0.032$ & $9.317\pm0.028$ & $0.729<\log{M}\leq0.993$ \\
$-1.442\pm0.037$ & $8.906\pm0.041$ & $0.993<\log{M}\leq1.257$ \\
$-1.062\pm0.043$ & $8.428\pm0.059$ & $1.257<\log{M}\leq1.522$ \\
$-0.833\pm0.053$ & $8.078\pm0.087$ &       $\log{M}>1.522$    \\
\hline
\multicolumn{3}{c}{$Z=0.0024$} \\
\hline
$-3.100\pm0.035$ & $9.881\pm0.004$ &       $\log{M}\leq0.245$ \\
$-2.479\pm0.033$ & $9.728\pm0.013$ & $0.245<\log{M}\leq0.551$ \\
$-2.128\pm0.035$ & $9.535\pm0.025$ & $0.551<\log{M}\leq0.857$ \\
$-1.649\pm0.040$ & $9.124\pm0.040$ & $0.857<\log{M}\leq1.163$ \\
$-1.156\pm0.047$ & $8.552\pm0.061$ & $1.163<\log{M}\leq1.469$ \\
$-0.862\pm0.061$ & $8.119\pm0.097$ &       $\log{M}>1.469$    \\
\hline
\multicolumn{3}{c}{$Z=0.004$} \\
\hline
$-3.207\pm0.045$ & $9.904\pm0.004$ &       $\log{M}\leq0.207$ \\
$-2.470\pm0.041$ & $9.751\pm0.014$ & $0.207<\log{M}\leq0.465$ \\
$-2.398\pm0.043$ & $9.718\pm0.025$ & $0.465<\log{M}\leq0.723$ \\
$-1.901\pm0.047$ & $9.358\pm0.040$ & $0.723<\log{M}\leq0.982$ \\
$-1.559\pm0.053$ & $9.023\pm0.059$ & $0.982<\log{M}\leq1.240$ \\
$-1.083\pm0.062$ & $8.432\pm0.084$ & $1.240<\log{M}\leq1.499$ \\
$-0.846\pm0.078$ & $8.076\pm0.126$ &       $\log{M}>1.499$    \\
\hline
\multicolumn{3}{c}{$Z=0.008$} \\
\hline
$-3.461\pm0.008$ &         $9.976\pm0.012$ &       $\log{M}\leq0.179$ \\
$-2.347\pm0.007$ &         $9.776\pm0.011$ & $0.179<\log{M}\leq0.404$ \\
$-2.727\pm0.008$ &         $9.930\pm0.011$ & $0.404<\log{M}\leq0.628$ \\
$-2.154\pm0.008$ &         $9.570\pm0.014$ & $0.628<\log{M}\leq0.853$ \\
$-1.848\pm0.009$ &         $9.309\pm0.015$ & $0.853<\log{M}\leq1.077$ \\
$-1.398\pm0.010$ &         $8.825\pm0.015$ & $1.077<\log{M}\leq1.302$ \\
$-1.134\pm0.012$ &         $8.451\pm0.017$ & $1.302<\log{M}\leq1.526$ \\
$-0.681\pm0.015$ &         $7.790\pm0.021$ &       $\log{M}>1.526$    \\
\hline
\end{tabular}
\end{table}

%TABLE A2 continued
\begin{table}
\contcaption{}
\begin{tabular}{ccr}
\hline \hline
\multicolumn{3}{c}{$Z=0.015$} \\
\hline
$-3.062\pm0.007$ & \llap{1}$0.096\pm0.011$ &       $\log{M}\leq0.237$ \\
$-2.425\pm0.007$ &         $9.945\pm0.010$ & $0.237<\log{M}\leq0.455$ \\
$-2.824\pm0.007$ & \llap{1}$0.127\pm0.010$ & $0.455<\log{M}\leq0.674$ \\
$-2.316\pm0.007$ &         $9.786\pm0.011$ & $0.674<\log{M}\leq0.892$ \\
$-1.926\pm0.008$ &         $9.438\pm0.012$ & $0.892<\log{M}\leq1.110$ \\
$-1.399\pm0.009$ &         $8.853\pm0.013$ & $1.110<\log{M}\leq1.328$ \\
$-1.180\pm0.063$ &         $8.562\pm0.067$ & $1.328<\log{M}\leq1.546$ \\
$-0.625\pm0.013$ &         $7.704\pm0.018$ &       $\log{M}>1.546$    \\
\hline
\end{tabular}
\end{table}

% TABLE A3
\begin{table}
\caption[]{Relation between the relative pulsation duration and birth mass,
$\log(\delta t/t)=D+
\Sigma_{i=1}^3a _i\exp\left[-(\log M [{\rm M} _\odot]-b_i)^2/2c_i^2\right]$.}
\begin{tabular}{ccccc}
\hline\hline
$D$    & $i$ & $a$   & $b$   & $c$   \\
\hline
\multicolumn{5}{c}{$Z=0.0012$} \\
\hline
$-5.2$ &  1  & 1.955 & 0.173 & 0.074 \\
       &  2  & 2.545 & 0.545 & 0.196 \\
       &  3  & 2.936 & 1.539 & 0.429 \\
\hline
\multicolumn{5}{c}{$Z=0.0015$} \\
\hline
$-4.9$ &  1  & 0.850 & 0.202 & 0.080 \\
       &  2  & 2.265 & 0.547 & 0.188 \\
       &  3  & 2.748 & 1.483 & 0.410 \\
\hline
\multicolumn{5}{c}{$Z=0.0019$} \\
\hline
$-4.9$ &  1  & 1.02 & 0.23 & 0.056 \\
       &  2  & 1.43 & 0.53 & 0.129 \\
       &  3  & 2.52 & 1.29 & 0.576 \\
\hline
\multicolumn{5}{c}{$Z=0.0024$} \\
\hline
$-4.8$ &  1  & 1.267 & 0.222 & 0.064 \\
       &  2  & 1.831 & 0.529 & 0.147 \\
       &  3  & 2.567 & 1.302 & 0.427 \\

\hline
\multicolumn{5}{c}{$Z=0.004$} \\
\hline
$-5.2$ &  1  & 1.510 & 0.229 & 0.065 \\
       &  2  & 1.702 & 0.508 & 0.149 \\
       &  3  & 3.258 & 1.320 & 0.519 \\
\hline
\multicolumn{5}{c}{$Z=0.008$} \\
\hline
$-3.96$ & 1 & 2.34 & 1.281 & 0.378 \\
        & 2 & 1.32 & 0.460 & 0.165 \\
        & 3 & 0.38 & 0.145 & 0.067 \\
\hline
\multicolumn{5}{c}{$Z=0.015$} \\
\hline
$-3.63$ & 1 & 2.18 & 1.238 & 0.238 \\
        & 2 & 1.36 & 0.514 & 0.127 \\
        & 3 & 0.30 & 0.228 & 0.084 \\
\hline
\end{tabular}
\end{table}

%==============================================================================
\section*{Acknowledgments}
We thank the staff at UKIRT for their excellent support of this programme. JvL
thanks the School of Astronomy at IPM, Tehran, for their hospitality during
his visits. We are grateful for financial support by The Leverhulme Trust
under grant No.\ RF/4/RFG/2007/0297, by the Royal Astronomical Society, and by
the Royal Society under grant No.\ IE130487. FST acknowledges financial
support from the Spanish Ministry of Economy and Competitiveness (MINECO)
under grant number AYA2013-41243-P. Finally, we thank the referee for her/his 
constructive report which prompted us to improve the manuscript. This work
has made use of BaSTI web tools.

%==============================================================================

\label{lastpage}
\end{document}